\documentclass[letterpaper,twocolumn,10pt]{article}
\usepackage{usenix-2020-09}

\usepackage{amsthm}

\newtheorem{theorem}{Theorem}[section]

\usepackage{hyperref}
\usepackage[normalem]{ulem}
\usepackage{xcolor}
\usepackage{amsmath,amsfonts}
\usepackage{graphicx}
\usepackage{textcomp}
\usepackage{enumitem}
\usepackage{listings}
\lstdefinestyle{mystyle}{
    language=Python,
    basicstyle=\ttfamily\footnotesize,
    keywordstyle=\color{blue!70!black}\bfseries,
    commentstyle=\color{gray!60!black}\itshape,
    stringstyle=\color{orange!80!black},
    numbers=left,
    numberstyle=\tiny\color{gray},
    stepnumber=1,
    numbersep=8pt,
    frame=single,
    backgroundcolor=\color{gray!10},
    breaklines=true,
    breakatwhitespace=true,
    showstringspaces=false,
    tabsize=4,
    captionpos=b
}

\usepackage{booktabs,siunitx,threeparttable}
\usepackage{tabularx}

\usepackage{subcaption}
\usepackage{soul}
\usepackage{comment}
\usepackage{xspace}
\usepackage{array}
\hypersetup{
	unicode=true,
        bookmarksnumbered=true,
	colorlinks=true,
	linkcolor={red!70!black},
	citecolor={red!70!black},
	urlcolor={blue!70!black},
	pdfborder={0 0 0}
}
\usepackage[capitalize,nameinlink]{cleveref}
\usepackage{flushend}
\usepackage{pifont}
\usepackage{threeparttable} % Add this to your preamble
\usepackage{multirow}
\newcommand{\eg}{e.g.,\xspace}

\newcommand{\ie}{i.e.,\xspace}

\newcommand*{\rom}[1]{\uppercase\expandafter{\romannumeral #1\relax}}

\newif\ifshowcomment
\showcommenttrue
\ifshowcomment
    \newcommand{\pantea}[1]{{\color{blue}[PZ: #1]}}
\else
    \newcommand{\pantea}[1]{}
\fi

\newcommand{\sysname}{{{Octopus}}\xspace}
\newcommand{\baseline}{{{FC}}\xspace}
\newcommand{\azure}{Azure\xspace}

\newcommand{\mpd}{{{MPD}}\xspace}
\newcommand{\mpds}{{{MPDs}}\xspace}

\newenvironment{denseitemize}{
\begin{itemize}[topsep=2pt, partopsep=0pt, leftmargin=1.5em]
  \setlength{\itemsep}{2pt}
  \setlength{\parskip}{0pt}
  \setlength{\parsep}{0pt}
}{\end{itemize}}

\newenvironment{denseenum}{
\begin{enumerate}[topsep=2pt, partopsep=0pt, leftmargin=1.5em]
  \setlength{\itemsep}{2pt}
  \setlength{\parskip}{0pt}
  \setlength{\parsep}{0pt}
}{\end{enumerate}}

\usepackage{titlesec}
\titlespacing*{\paragraph}{0pt}{0.5em}{0.5em}  

\usepackage{tikz}
\newcommand{\colcircnum}[2]{%
  \tikz[baseline=(n.base)]\node[draw,circle,fill=#1,inner sep=0.5pt](n){\textcolor{black}{#2}};%
}

\usepackage{xurl}   % better URL line breaking

\newcommand{\us}{\textmu{}s\xspace}

\usepackage{wrapfig}
\usepackage{booktabs}
\usepackage[table]{xcolor}

\usepackage{amsthm}

\newcommand{\yuhong}[1]{\textbf{\textcolor{purple}{Yuhong: #1}}}

\makeatletter
\renewcommand{\@maketitle}{%
  \newpage
  \null
  \vskip -1em               % reduce space above title
  \begin{center}%
    {\Large \bf \@title \par}%
    \vskip 2ex              % space between title and authors
    {\normalsize
      \begin{tabular}[t]{c}
        \@author
      \end{tabular}\par}%
    \vskip 7ex            % space below author block (reduce this!)
  \end{center}%
}
\makeatother

\begin{document}

%\title{Octopus: Low-Cost CXL Memory Pooling that Scales}
%\title{\sysname: Scalable Low-Cost CXL Memory Pooling}
\title{\sysname: Enhancing CXL Memory Pods via Sparse Topology}

%\vspace{10pt}

% \author{
% {\rm Yuhong Zhong}\\
% Columbia University
% \and
% {\rm Fiodar Kazhamiaka}\\
% Microsoft Azure
% \and
% {\rm Pantea Zardoshti}\\
% Microsoft Azure
% \and
% {\rm Shuwei Teng}\\
% Microsoft Azure\\[0.8em]
% \and
% {\rm Rodrigo Fonseca}\\
% Microsoft Azure
% \and
% {\rm Mark D. Hill}\\
% University of Wisconsin–Madison
% \and
% {\rm Daniel S. Berger}\\
% Microsoft Azure \& University of Washington
% }

\newcommand*{\cmu}{\includegraphics[scale=0.018]{figures/cmu.jpg}}%
\newcommand*{\intelicon}{\includegraphics[height=1em]{figures/intelicon.pdf}}%
\newcommand*{\microsoft}{\includegraphics[width=1em]{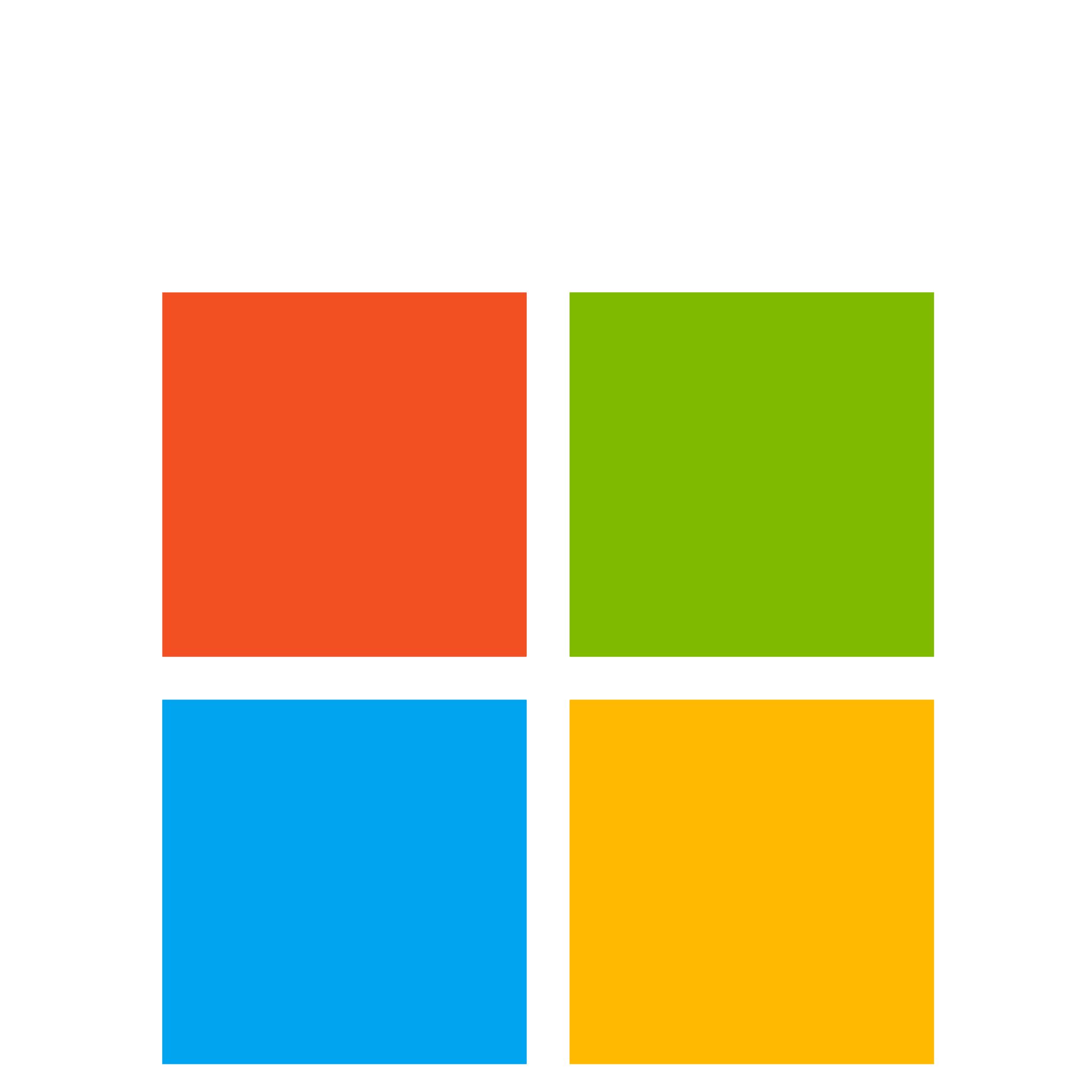}}%
\newcommand*{\uw}{\includegraphics[width=1em]{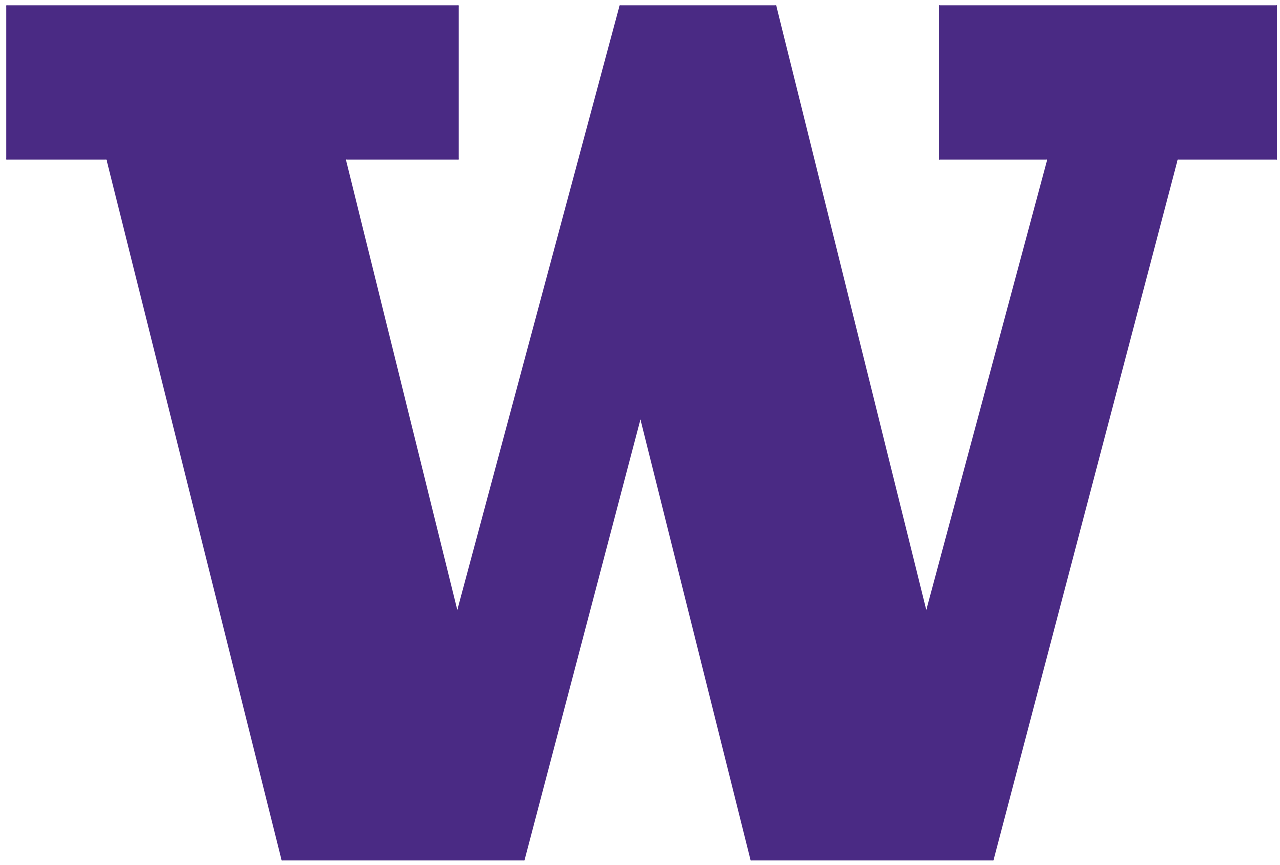}}%
\newcommand*{\columbia}{\includegraphics[width=1em]{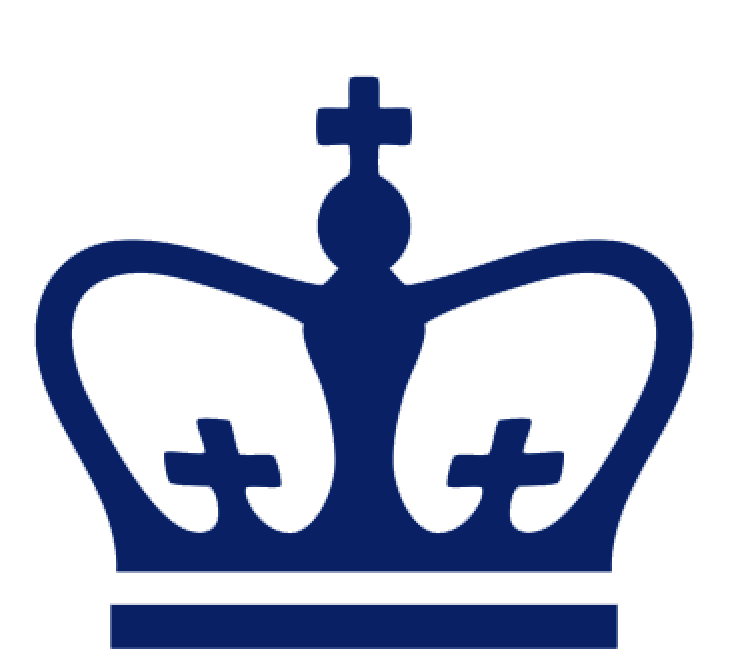}}%
\newcommand*{\umich}{\includegraphics[width=1em]{figures/umich.png}}%
\newcommand*{\uwmadison}{\includegraphics[width=1em]{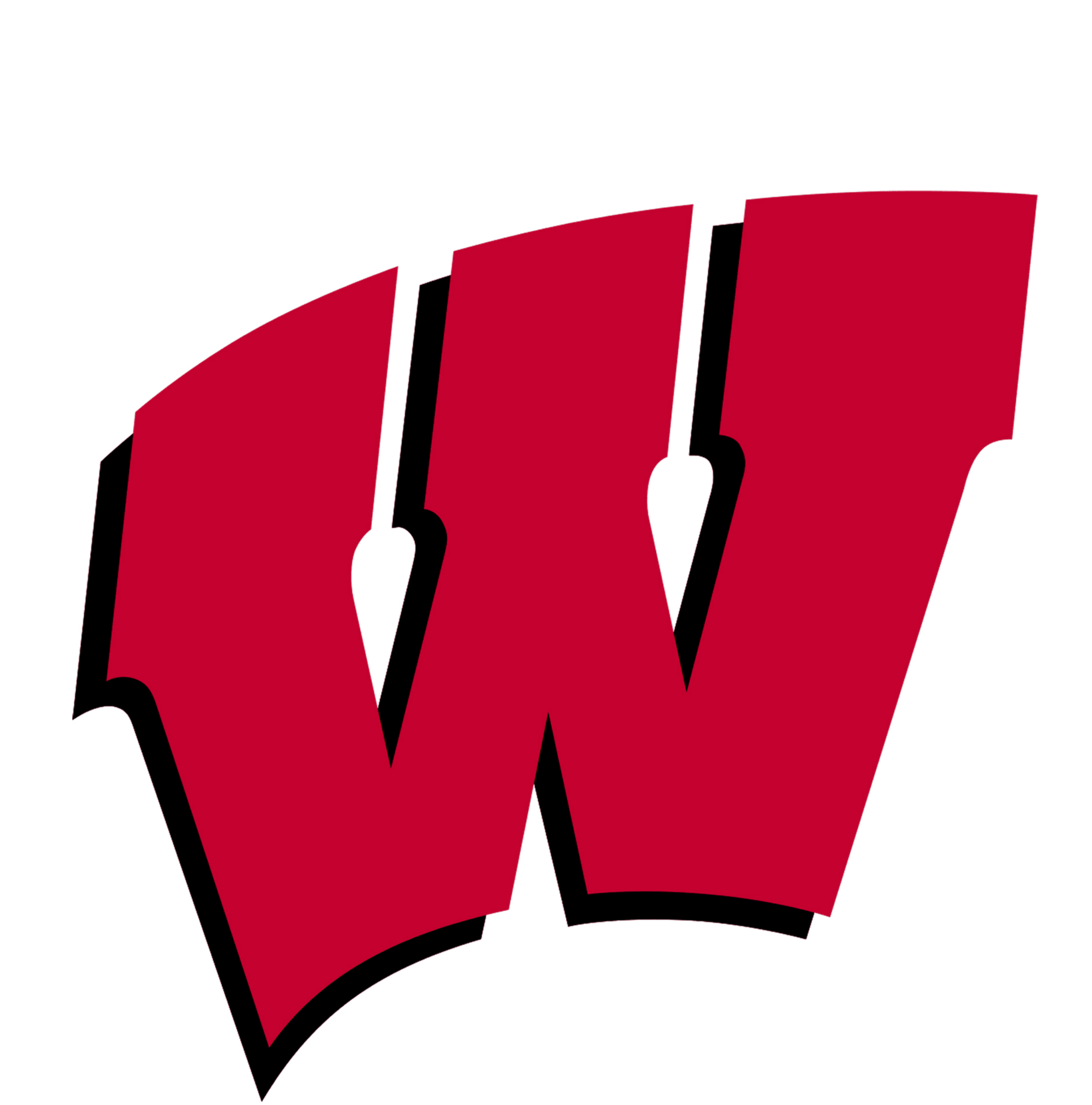}}%

\author{
    \large
    \rm Yuhong Zhong\,\columbia \quad
    Fiodar Kazhamiaka\,\microsoft \quad
    Pantea Zardoshti\,\microsoft \quad
    Shuwei Teng\,\microsoft \\
    % Rodrigo Fonseca\,\microsoft
    \large
    Rodrigo Fonseca\,\microsoft \quad
    Mark D. Hill\,\uwmadison \quad
    Daniel S. Berger\,\microsoft\,\uw\\[0.3em]
    {\it
    \columbia\,Columbia University \hspace{0.1in}
    \microsoft\,Microsoft Azure \hspace{0.1in}
    \uwmadison\,University of Wisconsin–Madison \hspace{0.1in}
    \uw\,University of Washington
    }
}

\begin{comment}
\begin{CCSXML}
<ccs2012>
   <concept>
       <concept_id>10010520.10010521.10010537.10003100</concept_id>
       <concept_desc>Computer systems organization~Cloud computing</concept_desc>
       <concept_significance>300</concept_significance>
       </concept>
 </ccs2012>
\end{CCSXML}

\ccsdesc[500]{Computer systems organization~Cloud computing}
\keywords{Cloud Architecture, Memory Pooling, CXL}
%\renewcommand{\shortauthors}{B. Reidys, P. Zardoshti, I. Goiri, et al.}
\renewcommand{\shortauthors}{Berger et al.}
%\renewcommand{\authors}{Benjamin Reidys, Pantea Zardoshti, Íñigo Goiri, Celine Irvene, Daniel Berger, Haoran Ma, Kapil Arya, Eli Cortez, Taylor Stark, Eugene Bak, Mehmet Iyigun, Stanko Novaković, Lisa Hsu, Karel Trueba, Abhisek Pan, Chetan Bansal, Saravan Rajmohan, Jian Huang, and Ricardo Bianchini.}
\end{comment}

%\thispagestyle{plain}
%\pagestyle{plain}

\maketitle
\vspace*{-4em}
% \footnotetext{Acadia is known for its unique hydrological system, where ponds are sparsely connected through streams and wetlands, forming a cohesive yet sparse network.}

%\pagenumbering{gobble}
\begin{comment}

\begin{abstract}
Compute Express Link (CXL) enables memory pooling across hosts to reduce costs and improve efficiency. However, existing CXL memory pool designs rely on exotic large pooling devices or expensive switches.

We propose \sysname, a cost-effective CXL pod design that uses near-commodity small pooling devices. \sysname eschews fully-connecting all hosts and pooling devices which enables scaling with these small devices and effectively reduces cost and access latency.

Simulations on production traces show \sysname achieves memory savings comparable to expensive pool designs. Hardware experiments confirm that \sysname reduces RPC latency by $3\times$ compared to RDMA and enables larger shuffle workloads. Our work formalizes \sysname topologies, develops memory allocation algorithms, and evaluates performance trade-offs through simulation and hardware testing.
\end{abstract}
    
\end{comment}

\begin{abstract}
The Compute Express Link (CXL) interconnect enables compute ``pods'' that pool memory across servers to reduce cost and improve efficiency.
These pods also facilitate pairwise communication whose needs conflict with pooling. Importantly,
% However, 
existing pod designs are small or require indirection through expensive switches.
These conventional designs implicitly assume that pods must fully connect all servers to all CXL pooling devices.

This paper breaks with this conventional wisdom by introducing \emph{\sysname} pods.
\sysname directly connects servers to low-port-count CXL pooling devices (\eg 4 ports) yet scales to large pods without switches by constructing a \emph{sparse} CXL topology in which each pooling device connects to a carefully chosen subset of servers.
\sysname explicitly balances ``overlap'', where two servers connect to the same pooling device: overlap reduces pooling efficiency but enables low-latency communication. \sysname resolves this tension by grouping servers into ``islands'' with low-latency intra-island communication and interconnecting islands to favor pooling.

% This paper breaks with this conventional wisdom to create \emph{\sysname} pods.
% \sysname directly connects servers to CXL pooling devices with few ports (\eg 4), but enables large pods without switches by connecting each pooling device to a carefully chosen subset of servers.
% \sysname's CXL interconnect balances ``collisions'', where two servers connect to the same a pooling device: collisions reduce pooling efficiency but enable low-latency communication.
% \sysname resolves this tension by grouping servers into islands with strong intra-island communication and interconnecting islands to favor pooling.
% \sysname connects each server to a bounded number of small multi-port pooling devices (\eg 8), and each pooling device connects to different subsets of servers, forming a sparse topology of servers and pooling devices.
% Despite no longer having a global memory pool,
% %\sysname might seem like a poor design as its memory pool is no longer a global resource. However, 
% we show that \sysname pods still effectively support memory pooling and useful communication and data sharing patterns. Relative to conventional pods, \sysname is more cost-effective (using near-commodity pooling devices) and enables larger pods (allowing greater pooling savings and greater communication reach).

We build a three-server CXL pod prototype and simulate scaled pods with 96 servers under measured device characteristics and physical constraints (1.5\,m copper cables).
On hardware, \sysname RPCs are 3.2$\times$ faster than in-rack RDMA and 2.4$\times$ faster than CXL switches.
In simulation, \sysname achieves net server cost savings of 3--5.4\% whereas CXL switches result in a net cost increase.
\end{abstract}

% 3%, 5.4% reduction

%\pagestyle{empty}

\section{Introduction}

\paragraph{Cloud memory pressure.}
General-purpose cloud systems face a growing memory shortfall.
Compute scales with die area ($n^2$), but the DRAM a socket can attach over DDR4/5 is “beachfront”-limited ($4n$).
As core counts and per-core performance have driven per-socket compute up by $5$--$10\times$ over the past decade, socket-attached DRAM capacity has grown only $2$--$3\times$~\cite{mutlu2013memory,lee2022dram,ewais2023disaggregated}.
Right-sizing cores or pushing more work to scale-out helps, but the most cost-effective lever is: \emph{attach more memory to the same socket}.

CXL memory expansion is the first step.
Compute Express Link (CXL) provides a practical path today to add capacity with CPU load/store semantics~\cite{das2024introduction}.
CXL-attached DRAM (“expansion”) is accessed like local memory (\texttt{ld/st}); it is supported by major CPU platforms~\cite{intel2024flat,amdcxl,armcxl} and shipping servers that double memory capacity~\cite{Stevescargall2025_CXLServerGuide,supermicrocxl,Dell2025_CXLPowerEdge,Lenovo2025_ThinkSystemV4_CXL,PenguinSolutions2025}.
While expansion adds latency ($\approx\!230\,\mathrm{ns}$ vs. $115\,\mathrm{ns}$), many web, key-value, analytics, and database workloads tolerate this NUMA-like penalty~\cite{zhong2024managing,liu2025cxl,ahn2024examination}.

\begin{figure}
    \centering
    % \begin{subfigure}{0.23\textwidth}
    % \includegraphics[clip,width=0.99\textwidth]{figures/interisland}
    % \caption{Inter-island connectivity.}
    % \label{fig:interisland}
    % \end{subfigure}
    % \begin{subfigure}{0.23\textwidth}
    %     \centering
    %     \includegraphics[width=0.99\textwidth]{figures/interisland_highlevel}
    %     \caption{96-servers / six islands.}
    %     \label{fig:sixisland}
    % \end{subfigure}
    % \includegraphics[width=0.38\textwidth]{figures/island-interconnect.pdf}
    \includegraphics[width=\columnwidth,trim=0.5cm 0 0 0,
                 clip]{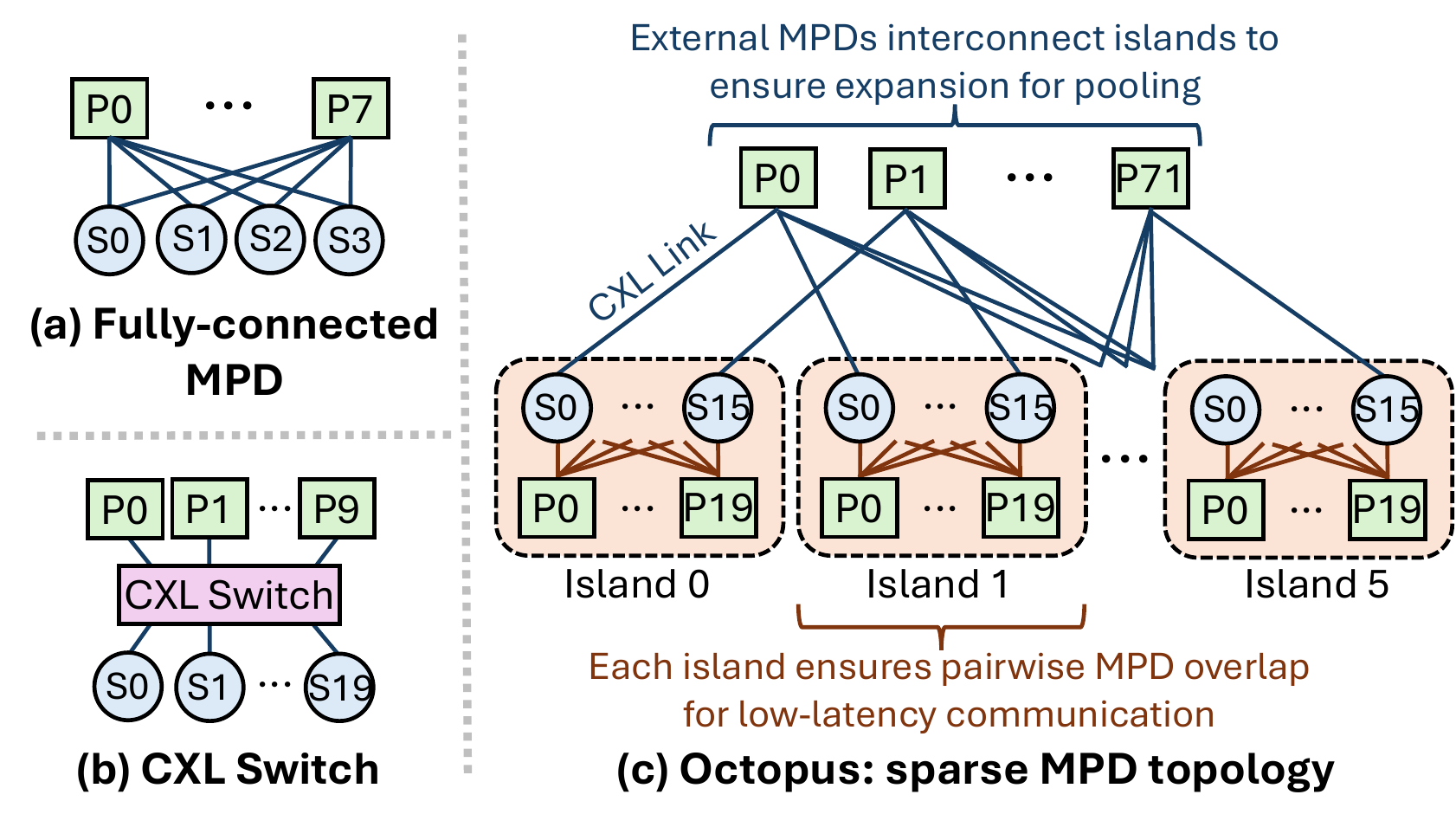}
    \vspace*{-2em}
    \caption{High-level view of \textbf{(a)} a fully-connected \mpd pod, \textbf{(b)} a CXL switch pod, and \textbf{(c)} an \sysname pod with six islands (96 servers (``S'') and 192 \mpds (``P'')).
    Within each \sysname island, servers use a subset of CXL ports to connect to island-specific \mpds, guaranteeing pairwise \mpd overlap for low-latency communication. Remaining server ports connect to external \mpds that interconnect islands and enable effective memory pooling.
    % High-level view of an \sysname topology with multiple islands.
    % (a) shows how $X_i$ host ports are allocated to internal \mpds to form an island for data sharing.
    % The remaining host ports (e.g., 3) connect to external \mpds that interconnect islands for pooling.
    % (b) shows a six-island \sysname topology with 96 servers at a high level. The numbers show the number of \mpds that each pair of islands share, we optimize for symmetry but this is not always mathematically possible. \yuhong{better explain how islands are interconnected.}
    }
    \vspace*{-1em}
    \label{fig:interconnect-island}
\end{figure}

% \begin{figure}
%     \centering
%     \begin{subfigure}{0.17\textwidth}
%         \includegraphics[clip,width=0.99\textwidth]{figures/fully-connected-mpd.pdf}
%         \caption{Inter-island connectivity.}
%         \label{fig:interisland}
%     \end{subfigure}
%     \begin{subfigure}{0.23\textwidth}
%         \includegraphics[clip,width=0.99\textwidth]{figures/switch-topo.pdf}
%         \caption{Inter-island connectivity.}
%         \label{fig:interisland}
%     \end{subfigure}\\
%     \begin{subfigure}{0.38\textwidth}
%         \centering
%         \includegraphics[width=0.99\textwidth]{figures/island-interconnect.pdf}
%         \caption{High-level view of an \sysname pod with six islands (96 servers (``S'') and 192 \mpds (``P'')).
%     Within each island, servers use a subset of CXL ports to connect to island-specific \mpds, guaranteeing pairwise \mpd collisions for low-latency communication. Remaining server ports connect to external \mpds that interconnect islands and enable memory pooling.}
%         \label{fig:interconnect-island}
%     \end{subfigure}
    
%     % \includegraphics[width=0.38\textwidth]{figures/island-interconnect.pdf}
%     \vspace*{-1em}
%     \caption{Overview
%     }
%     \vspace*{-1em}
%     \label{fig:topo-overview}
% \end{figure}

This paper explores CXL ``pods'' spanning multiple servers with CXL as a scale-up interconnect.
Pods enable \emph{memory pooling}, which reduces DRAM spend (often half of server cost) by multiplexing capacity demand across servers~\cite{pond:asplos2023,wahlgren2022evaluating,lee2023elastic,ahn2024examination,xu2024faasmem,ewais2023disaggregated}.
The intra-pod CXL fabric also creates a \emph{low-latency communication and data sharing} fabric that improves performance of several workload classes.
Specifically, pod-wide RPCs, coordination, and collective-based data movement can be sped up by sharing memory buffers over the CXL fabric~\cite{rpcool,zhang2024dmrpc,baumstark2024so,cxl-shm,suetterlein2024synchronization,ahn2024mpi,cmpi,zhong2025oasis,zhong2025beware}.
Realizing these promises, however, largely depends on the CXL fabric's performance and topology.

\begin{comment}
This paper explores the next step of creating a CXL "pod" with multiple hosts and memory modules with CXL as a "scale-up" interconnect.
CXL pods have two distinct advantages, 1) enables intra-pod pooling of memory capacity and 2) low-latency intra-pod data exchange that bypasses the network stack.
As with GPU pods, we expect inter-pod communication to still use Ethernet as its scale-out network.
Key CXL pod use cases should improve performance-per-dollar to offset the cost of the CXL fabric:
\emph{(i)} \emph{capacity pooling} to multiplex demand and reduce DRAM spend (often half of server cost)~\cite{pond:asplos2023,wahlgren2022evaluating,lee2023elastic,ahn2024examination,xu2024faasmem,ewais2023disaggregated};
\emph{(ii)} \emph{pod-wide RPC and coordination} via shared buffers;
and \emph{(iii)} \emph{collective-based} data movement~\cite{ahn2024mpi}.
\end{comment}

% \begin{figure}[h]
%     \centering
%     \includegraphics[height=1.45in]{figures/designfig}
%     \caption{\sysname combines offering low latency for pooling memory and communication use cases, while offering a large pooling domain for multiplexing memory demands.
%     \yuhong{fix the diagram.}}
%     \label{fig:introfig}
% \end{figure}

\paragraph{Existing pod options and their tradeoffs.}
Two practical ways exist to connect servers to CXL memory modules:
\begin{denseitemize}
\item \emph{Multi-ported devices (MPDs)} collapse a tiny on-device interconnect with the memory controllers so multiple servers can attach directly, \emph{without a CXL switch}~\cite{pond:asplos2023,wagh2021cxl,seagate2024cxl,arm2024cxl,ha2023dynamic,mackey2024cxl,hyatt2023quest}.
Today's MPDs are small (2--4 ports; $\approx\!270$\,ns access)~\cite{leo2023,marvell2024structera,hyperscale2023cxl,huang2025txcocket,seagate_cma,seagate2024cxl,weisgut2025cxl}, and thus thought to limit pods to a handful of servers. 
Prior work assumes \emph{fully-connected} pods, in which every \mpd connects to every server, to enable hardware interleaving across \mpds for higher bandwidth~\cite{pond:asplos2023,wagh2021cxl,seagate2024cxl,arm2024cxl,ha2023dynamic,mackey2024cxl}.
\item \emph{CXL switches} expand connectivity (dozens of ports), but add (de)serialization hops, raising latency and server capital expenditure (CapEx)~\cite{das2024introduction,berger2023design,levis2023case}.
\end{denseitemize}
However, both options have key drawbacks: fully-connected \mpds limit pooling benefits due to their small pod sizes, whereas CXL switches increase latency and CapEx to the point where switch costs can outweigh the pooling savings.

% In practice, three pressures collide: (1) build a pod large enough to pool effectively; (2) keep common access/communication paths short and diverse enough for other use cases; and (3) keep pod CapEx low enough that the savings exceed the cost of the fabric.

\paragraph{Our approach: sparse \mpd topologies.}
We address the pod size limitations of \mpd topologies by breaking from prior work and considering \emph{sparsely connected} servers and \mpds.
% We keep the simplicity of \mpd-based pods but break from prior work by considering \emph{sparsely-connected} servers and \mpds.
In a sparsely-connected topology, each server connects to multiple \mpds, which in turn connect to \emph{different sets of servers}.
This allows pods to scale beyond the \mpd port count, \eg dozens of servers with 4-port \mpds, without the expense and latency of CXL switches~\cite{levis2023case}.
% \yuhong{need visualization.}

We propose \emph{\sysname}, a deployable, cost-conscious CXL pod design built on sparse \mpd topologies that targets both memory pooling and low-latency communication.
\sysname accounts for practical CXL constraints, including the length of copper CXL cables, port count limits of commodity \mpds, and the number of CXL ports available on modern servers.

% However, designing sparse \mpd topologies that simultaneously support both use cases is challenging.
% Low-latency communication and memory pooling favor conflicting topology properties: low-latency communication benefits when a pair of servers \emph{collides} on a common \mpd to directly share memory buffers, whereas memory pooling prefers avoiding \mpd collisions among sets of "hot" servers so that their excess memory demand can be spread across more \mpds~(\S\ref{sec:tension}).

However, designing sparse \mpd topologies that simultaneously support both use cases is challenging.
Low-latency communication and memory pooling have conflicting requirements in terms of \emph{\mpd overlap}, \ie whether servers connect to a common \mpd.
Low-latency communication favors pairwise \mpd overlap between any two servers, whereas memory pooling disfavors \mpd overlap among "hot" servers so that their excess memory demand can be spread across as many \mpds as possible~(\S\ref{sec:tension}).

To overcome this challenge, we observe that low-latency communication is typically needed only at modest scale (\eg one dozen of servers), whereas effective memory pooling requires much larger scale (dozens to hundreds of servers).
\sysname therefore organizes each pod into multiple \emph{islands}~(Figure~\ref{fig:interconnect-island}). Within each island, servers use a subset of ports to connect to island-specific \mpds arranged using Balanced Incomplete Block Design (BIBD)~\cite{pignolet-bibd,cypher2015configuring}, which guarantees pairwise \mpd overlap for low-latency communication. The remaining server ports are used for interconnecting islands via additional \mpds, limiting \mpd overlap among hot servers and enabling effective memory pooling across the pod~(\S\ref{sec:logicaltopo}).

We implement a minimal prototype of \sysname on real hardware spanning three servers and three \mpds.
\sysname closely matches the performance of existing CXL memory expansion systems, where a CXL switch would otherwise increase the slowdown of 25\% of applications by 3$\times$. 
\sysname's communication latency is 3.2$\times$ lower than in-rack RDMA, 2.4$\times$ lower than a CXL switch, and 4.5$\times$ lower than other sparse \mpd topologies with good memory pooling savings. 

We simulate scaled-up \sysname topologies of 25, 64, and 96 servers based on 4-port \mpds.
\sysname achieves similarly low intra-island communication latency as fully-connected \mpd topologies.
% Figure~\ref{fig:introfig} shows how \sysname achieves similarly low latency to fully-connected MPD topologies from the CXL literature~\cite{pond:asplos2023}.
\sysname's savings from memory pooling alone can pay for the extra device costs both in cases with existing CXL expansion (5.4\% overall cost reduction) and without CXL in place (3.0\% overall cost reduction).
CXL switches always cost more, even after factoring in their pooling savings.
We open source \sysname pods and cost models at \url{https://github.com/yuhong-zhong/Octopus-CXL-Pod}.

While CXL can outperform Ethernet and RDMA, it does not replace them.
As with GPU pods, we expect inter-pod communication to still use Ethernet or RDMA.
Unlike GPU pods, however, CXL use cases are cost sensitive; our work shows how to make the intra-pod CXL fabric cost-effective across multiple use cases.

\paragraph{Contributions.} We make the following contributions:
\begin{denseitemize}
    % \item First paper to question the assumption that CXL pods must fully-connect servers and pooling devices.
    \item We propose \sysname, the first CXL pod design based on sparse \mpd topologies, challenging the assumption that CXL pods must fully connect servers and pooling devices.
    \item We systematize practical CXL pod constraints and identify key \mpd topology properties for CXL use cases.
    % \item Systematize practical CXL pod constraints and objectives. %Memory allocation and buffer placement algorithms for \sysname topologies
    % \item Show fully-connected is unnecessary for large CXL use cases
    \item We evaluate \sysname using real sparse \mpd hardware and simulation, and use a cost model to compare cost savings across CXL pod designs.
\end{denseitemize}

\begin{figure}[t]
  \centering
  \begin{minipage}[t]{0.5\linewidth}
    \vspace{0pt}
    \includegraphics[width=0.92\linewidth]{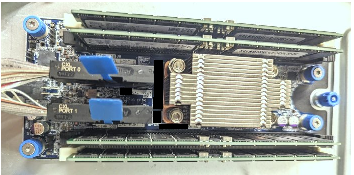}
  \end{minipage}%\hfill
  \begin{minipage}[t]{0.495\linewidth}
    \vspace{0pt}
    \centering\footnotesize
    \setlength{\tabcolsep}{4.5pt}\renewcommand{\arraystretch}{0.95}
    \begin{tabular}{@{}ll@{}}
      \toprule
      \textbf{Device} & \textbf{P50} \\
      \midrule
      CXL expansion      & 230--270 ns \\
      CXL 2/4-port MPD   & 260--300 ns \\
      CXL switch         & 490--600 ns \\
      RDMA via ToR           & 3550 ns \\
      \bottomrule
    \end{tabular}
  \end{minipage}
  \caption{\textbf{Left:} example two-port CXL device. \textbf{Right:} Load-to-use read latency (P50) on random 64-byte cachelines accessed via different CXL devices and RDMA on Intel Xeon 6 and AMD Turin.}
  \label{fig:cxldev-with-lat}
  \vspace*{-1em}
\end{figure}

\section{State of CXL in 2026}\label{sec:bg}

\begin{figure*}[t]
\begin{minipage}[t]{0.36\textwidth}
    \vspace{0pt}
  % Top: die-area sketch (roughly half-width to match the screenshot)
  \includegraphics[width=0.96\textwidth]{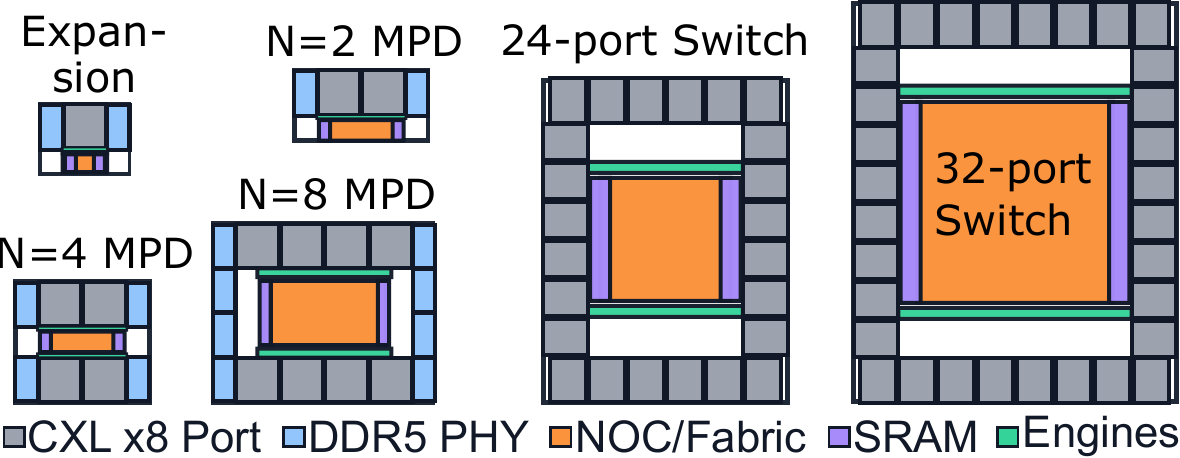}\\[-0.35em]
  \end{minipage}\hfill
  \begin{minipage}[t]{0.39\textwidth}
    \vspace{0pt}
    \footnotesize
    \setlength{\tabcolsep}{4.5pt}\renewcommand{\arraystretch}{0.95}
    \begin{threeparttable}
      \begin{tabular}{
        l
        S[table-format=2.0]  % CXL x8
        S[table-format=2.0]  % DDR5
        S[table-format=3.0]  % area
        S[table-format=4.0]  % price
      }
      \toprule
      {Type} & {CXL$\times$8} & {DDR5} & {Area [\si{\milli\metre\squared}]} & {Price [\$]} \\
      \midrule
      Expansion & 1  & 2  & 16  & 200 \\
      MPD       & 2  & 2  & 18  & 240 \\
      MPD       & 4  & 4  & 32  & 510 \\
      MPD       & 8  & 8  & 64  & 2650 \\
      Switch    & 24 & 0  & 120 & 5230 \\
      Switch    & 32 & 0  & 209 & 7400 \\
      \bottomrule
      \end{tabular}
    \end{threeparttable}
  \end{minipage}\hfill
  \begin{minipage}[t]{0.24\textwidth}
    \vspace{0pt}
    \footnotesize
    \setlength{\tabcolsep}{6pt}\renewcommand{\arraystretch}{0.95}
    \begin{threeparttable}
      \begin{tabular}{
        l % length
        l
        S[table-format=3.0]  % price
      }
      \toprule
      CXL Cable &  & \\
      {Length [m]} & {AWG} & {Price [\$]} \\
      \midrule
      0.50 & 30 & 23 \\
      0.75 & 28 & 29 \\
      1.00 & 28 & 36 \\
      1.25 & 26 & 55 \\
      1.50 & 26 & 75 \\
%      30 & OM4 & 800 \\
      \bottomrule
      \end{tabular}
    \end{threeparttable}
  \end{minipage}

  \vspace{-0.25em}
  \caption{Cost model for CXL devices and cables. \textbf{Left:} Die area estimates for devices with different CXL and DDR5 configurations. \textbf{Middle:}~Prices based on a yield and markup model. Die area figures are simplified but based on a real 8-port \mpd layout and modern IO dies. \textbf{Right:}~Cable prices are based on copper material prices and markup.}
  \label{fig:die-area-and-cost}
  \vspace*{-1em}
\end{figure*}

%\subsection{CXL Overview}

% This section describes how CPUs access CXL-attached memory, and how CXL can act as a scale-up intra-pod network. 

The Compute Express Link (CXL) interconnect standard itself is described in detail in a recent tutorial~\cite{das2024introduction}.
We discuss real devices and constraints when using CXL.mem today.
When a CPU issues a \texttt{ld} to an address associated with a CXL device and CPU caches do not have the data, the CPU sends a CXL.mem read \emph{flit} to the device.
CXL.mem flits use PCIe physical lanes and electrical signaling but implement custom low-latency protocol layers.
The device reads the cache line from its (DDR4 or DDR5) DRAM and sends it back to the CPU, which places it into processor caches and the core.

\paragraph{Latency and bandwidth.}
On Intel Xeon 6 platforms, a local DDR5 read takes about 115\,ns.
Reading from a good CXL.mem expansion device takes 200--300\,ns.
Using a CXL bus analyzer, we measure that 75--170\,ns of this latency is due to the CPU (most of the variability), 65\,ns is due to CPU port round-trips and flight time, 25\,ns is due to device-internal processing, and 35--40\,ns is due to DRAM access.
On AMD Turin platforms, performance is similar.
Devices typically offer multiple DDR4 or DDR5 DRAM channels.
%, random accesses behave like local DRAM.

%For example, a balanced-bandwidth configuration roughly matches a x8 (PCIe5) CXL port on the controller with two DDR4 channels, or one DDR5 channel, offering roughly 25-30 GB/s.
A single CPU typically has multiple CXL ports. %, and supports interleaving at 256B granularity across multiple EMCs. 
For example, Intel Xeon 6 production platforms use 64 CXL lanes per CPU socket, \eg ~\cite{lenovo2024thinksystem,asrock2024gnrd8,kalodrich2024supermicro,kaytus2024kr2280v3, kennedy2024lenovo}.
% example: https://www.servethehome.com/lenovo-has-a-cxl-memory-monster-astera-labs-intel-xeon/
% four socket, 16 DIMMs per socket == 64
% another 64 DIMMs via CXL, so 16 CXL cards, 4 per socket
A single $\times 8$ CXL port\footnote{A $\times8$ CXL port has 8 CXL lanes. Similarly, a $\times16$ port has 16 lanes.} offers 25--30\,GiB/s of read-only bandwidth, and twice that for $\times 16$.
The CPU socket's CXL lanes are configurable as four $\times 16$ CXL ports or eight $\times 8$ CXL ports.
In aggregate, a CPU socket can thus read from CXL devices at 200--240\,GiB/s.
%CXL is fully bidirectional; however, in practice devices may be underprovisioned on the DDR4/5 memory controller to serve fully bidirectionally.
%Firmware can interleave accesses across multiple CXL devices at 256B granularity, presenting a single NUMA node to the OS.

\paragraph{Device types.}
There are three CXL.mem device types:

\colcircnum{white}{1}
An \emph{expansion device} offers a single CXL port and exposes its memory to a single CPU.

\colcircnum{white}{2}
A \emph{multi-ported device (\mpd)} integrates $N$ CXL ports, allowing $N$ CPUs to connect to the same controller and access its memory concurrently.
Both expansion devices and \mpds are available today, and some products support both modes.
$N=2$ \mpds with two $\times 8$ ports (\eg Figure~\ref{fig:cxldev-with-lat}) are offered by AsteraLabs~\cite{leo2023}, Marvell~\cite{marvell2024structera}, Google and Meta~\cite{hyperscale2023cxl}, and xFusion~\cite{huang2025txcocket}; Seagate and SKH have $N=4$ prototypes~\cite{seagate_cma,seagate2024cxl,SKHynix2025,weisgut2025cxl}.
On our lab system, our $N=2$ MPD offers 267\,ns read latency and 28\,GiB/s read bandwidth per port.
%MPDs with 2 ports and 270 ns latency are available as near commodity today~\cite{leo2023,marvell2024structera,hyperscale2023cxl,huang2025txcocket}, and 4-port MPDs are available as prototypes~\cite{seagate_cma,seagate2024cxl,weisgut2025cxl}.
%MPDs could overcome CXL pooling's cost and latency challenges.

\colcircnum{white}{3} A \emph{CXL switch} offers up to 32 $\times 8$ CXL ports and can forward CXL packets between them, \eg the XC50256 switch offered by XConn~\cite{xconn2024switch}.
In today's CXL 2.0 standard, switches
connect to servers and expansion devices, but
cannot connect to other switches.
For every flit round-trip, a switch must deserialize and reserialize the flit twice, which adds at least 220\,ns of latency~\cite{xconn2024latency}~(Figure~\ref{fig:cxldev-with-lat}).
% with actual measurements somewhat higher.
This added latency also reduces effective bandwidth by increasing the bandwidth-delay product, as CPUs often support limited outstanding requests to fully utilize the link~\cite{cxl-melady,liu2026performance}.

\paragraph{CXL pods.} A \emph{CXL pod}~\cite{tigon,huang2025pasha} is a group of servers connected to shared CXL devices, enabling direct access to device-attached memory and supporting memory pooling and shared memory buffers across the pod. There are two common ways to build a CXL pod:
\colcircnum{white}{1} One approach uses \mpds, where each \mpd connects directly to several servers; prior \mpd-based pods assume a fully-connected design, in which every \mpd connects to every server within the pod~\cite{pond:asplos2023}. Using \mpds avoids expensive switches and keeps latency low, but limits pod size by the \mpd port count.
\colcircnum{white}{2} The other approach uses CXL switches~\cite{yang2025beluga,pond:asplos2023}, which fan out connectivity so many servers can connect to many devices; this supports larger pods but incurs higher latency and cost.

\paragraph{Physical constraints.}
CXL devices connect to the CPU via PCIe5 pins and cables, whose reach is limited by insertion loss.
At 16\,GHz, the total budget is 36\,dB; after the CPU package, motherboard, and \mpd board consume $\approx 26$\,dB, only $\approx 10$\,dB remains for the cable and connectors, \emph{constraining cable lengths to $\le$\,1.5\,m}~\cite{Samtec2021,Telian2023}.
Longer runs would require retimers~\cite{astera2024aries} or optical cables~\cite{samtec_pcuo}, both of which add latency, power, or cost.

% Servers can concurrently access the memory behind a \mpd or behind a switch.
% CXL 2.0 does not support hardware coherence, so software must ensure coherence~\cite{das2024introduction}.

% In your preamble:
% \usepackage{booktabs}
% \usepackage{siunitx}
% \sisetup{group-minimum-digits=4}

\section{Modeling CXL Device Cost}\label{sec:pdbg}

To understand the cost tradeoffs of different CXL pod designs, we model the cost of CXL devices and cables.
As is common with datacenter server components, vendor prices are subject to non-disclosure agreements~\cite{barroso2018datacenter}.
We thus rely on a die area model that captures typical IO-pad limitations~\cite{smith2005integrated}.

% We estimate die area and model the cost of different die area with four three data points.
% First, we use a 5~nm design for an \mpd with eight $\times 8$ ports, and 8 DDR5 channels proposed by ARM~\cite{arm2024cxl}.
% Second, we have scaled-up die area estimates from the 6~nm design of the AMD Zen4 (Genoa) IO Die~\cite{amd2023issc}, and we can estimate the die cost by interpolating the common factor of the IO die from the cost of CPUs with different core counts.
% Third, we have spoken to supply chain experts on pricing of expansion and 2-port devices today.
% We also model cable costs based on copper and assembly prices, as well as optical cable prices from a recent vendor demonstration~\cite{samtec_pcuo}.

We estimate die area and cost using three data points.
First, we use a 5\,nm design and its die area for an \mpd with eight $\times 8$ CXL ports and eight DDR5 channels, as proposed by ARM~\cite{arm2024cxl}.
Second, we scale die-area estimates from the 6\,nm AMD Zen 4 (Genoa) I/O die~\cite{amd2023issc} and infer die cost by interpolating the I/O-die contribution from CPUs with different core counts.
Third, we incorporate pricing for today's expansion devices, 2-port devices, and switches based on discussions with supply-chain experts and public numbers~\cite{yang2025beluga}.
In addition, we model cable costs using copper and assembly prices.
% , as well as optical cable pricing from a recent vendor demonstration~\cite{samtec_pcuo}.

\paragraph{\mpds.} Figure~\ref{fig:die-area-and-cost} shows our die area estimates, CXL device costs, and cable costs.
The cheapest device is a single-ported ($\times 8$) expansion device with two DDR5 channels, at \$200.
Compared to expansion devices, we provision \mpds with twice the number of CXL ports per DDR5 channel (\ie one $\times 8$ CXL port for each DDR5 channel), and assume this CXL-port-to-DDR-channel ratio for all \mpds.
This makes an $N=2$-port \mpd slightly more expensive than an expansion device, which is pessimistic because we know that in practice many expansion devices actually use one $\times 16$ that can be bifurcated into two $\times 8$ ports.
An $N=4$ \mpd amortizes some of the fixed area overheads, but we assume that vendors would charge a slightly higher markup.
At $N=8$, \mpds are IO-pad limited and prices increase significantly in our model.

\paragraph{Switches.} Our cost model shows that switches with 24 or 32 ports are substantially more expensive than \mpds: a 32-port switch would cost \$7,400.
A recent paper reports that the XConn XC50256, a 32 $\times 8$-port CXL switch, is priced at approximately \$5,800~\cite{yang2025beluga}.
This gap is largely explained by process technology: while our model assumes 5\,nm or 6\,nm nodes for consistency with modern CPUs, CXL switches are often fabricated on mature nodes such as TSMC 16\,nm to reduce wafer and NRE costs~\cite{xconn-cxl-slides}.
Even at 16\,nm, switches remain an order of magnitude more expensive than \mpds.

\paragraph{Power.}
\mpd-based CXL pods are also more power efficient than switch-based designs. Using a simple additive model where each CXL port consumes 2\,W and assuming $X=8$ CXL ports per server, \mpd pods incur about 72\,W per server versus 89.6\,W for switch-based pods (24\% more), due to the switch silicon and expansion devices. Although the per-server difference is modest (about 3\% of total power at 500\,W), the overhead compounds at scale.

\begin{comment}
\paragraph{\mpds vs. switches.} \yuhong{Seems redundant, maybe merge with the eval cost section?} Back-of-the-envelope cost calculations show why \mpds can be a cost-effective building block for CXL pods, while switches can be too expensive.
Say 96\,GB DDR5-5600 mainstream DIMMs cost \$432 (\$4.50\,GB as a common assumption~\cite{Hyatt2024}).
Each host uses 8 CXL expansion devices which are 32 DIMMs assuming two DIMMs per channel.
% $\$ 13.8k$ in DRAM cost.
Switching to $N=2$ \mpds increases CXL device cost overall by $\$ 320$, which less than the price of one DIMM.
Therefore, we would have to save at least 3\% in memory capacity by pooling to offset the extra cost.

In contrast, switches would have to save much more because they strictly add cost as we still need to buy the expansion devices.
For example, using a 24-port switch means using 12 ports for hosts and 12 ports for expansion devices.
Every host still needs eight $\times 8$ ports, to match bandwidth, which means multiple switches are needed.
Even just adding three switches exceeds the entire cost of the CXL-attached DRAM for a single host.
We remark that \$4.50\,GB is a low estimate; depending on the DIMM size and vendors, prices can go up significantly, especially for 128\,GB DIMMs~\cite{microndram,hpdram}.

%We also state approximate latency expectations that we again ground in anonymized device measurements.
%We later show that \textbf{XSmall} \mpds can be a cost-effective building blocks for CXL pods.
\end{comment}

\section{CXL Use Cases}\label{sec:usecases}

\begin{figure}[!t]
  \centering
  \includegraphics[width=0.91\linewidth]{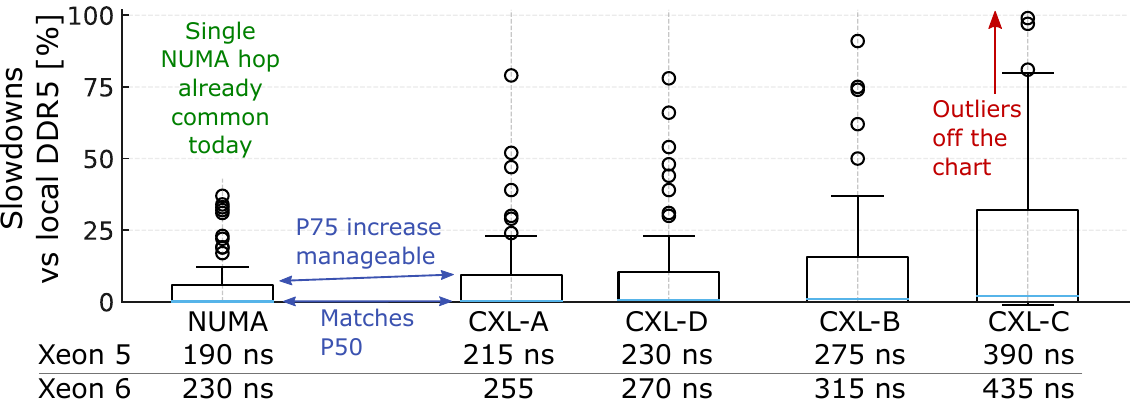}
  \caption{Box plots of workload slowdowns under different CXL latencies show that an increasing fraction of workloads sees slowdown around 390\,ns on Xeon 5, which is equivalent to 435\,ns on Xeon 6.}
  \label{fig:latency:sensitivity}
  \vspace*{-0.5em}
\end{figure}

\begin{figure}
    \centering
    \includegraphics[width=0.45\textwidth]{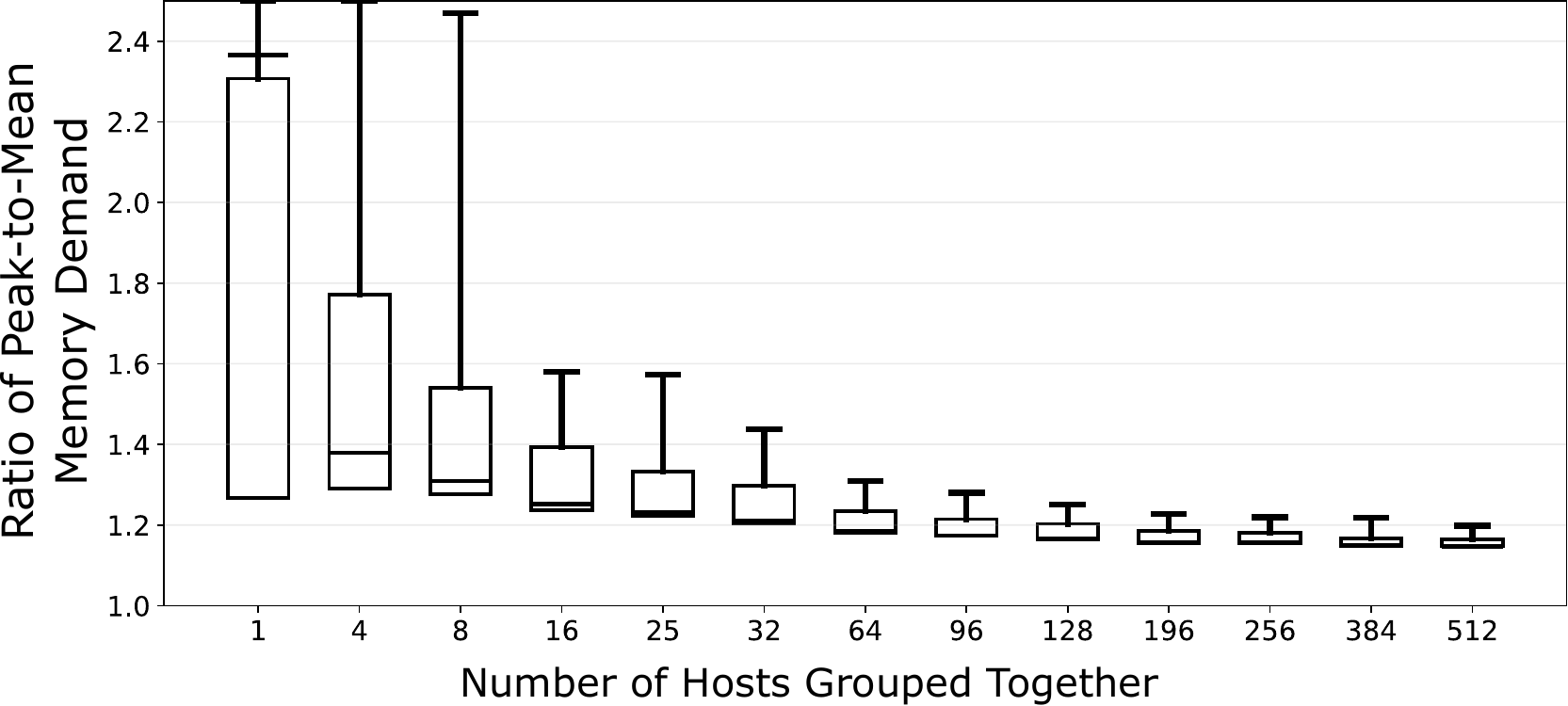}
    \vspace*{-0.5em}
    \caption{Ratio of peak memory demand to average memory demand across groups of servers of different sizes show that we need to group large numbers of servers to reduce outliers~\cite{liu2026performance}.
    Even groups of 25-32 servers would still need about 1.5$\times$ memory capacity to handle peaks.
    This data is based on VM traces from \azure, gathered from ten production clusters over two weeks~\cite{coach-asplos}.}
    \label{fig:poolingsavings}
    \vspace*{-1em}
\end{figure}

We briefly review CXL memory capacity expansion~(\S\ref{sec:use-case-expansion}), which is already deployed in production~\cite{azure-cxl-m-series,berger2026cxlpractice}, as well as memory pooling~(\S\ref{sec:use-case-pooling}) and shared memory use cases~(\S\ref{sec:use-case-sharing}).

\subsection{Memory Capacity Expansion}
\label{sec:use-case-expansion}

DDR5 memory capacity per unit socket performance has decreased over time, as core counts grow faster than the number of available DDR channels~\cite{berger2026cxlpractice}.
This is a structural trend: DDR and other parallel memory interfaces (LPDDR, HBM) consume large bundles of pins, routing layers, and on-package PHY area per channel, limiting how many channels fit on a package.
Individual DIMMs are also capacity-limited, and high-capacity parts carry steep \$/GiB premiums (\eg 3D stacking with TSV~\cite{Lenovo2025_TruDDR5_256GB,HPE2025_3DS256GB}).
CXL addresses both constraints by using serial links that deliver 5--10$\times$ more capacity per pin than DDR, decoupling the memory controller from the CPU package and enabling fan-out.
Practically, CXL can 2--2.5$\times$ socket memory capacity today, and CXL-based memory expansion is already deployed in production in \azure~\cite{azure-cxl-m-series,berger2026cxlpractice}.

CXL memory expansion is latency sensitive.
Figure~\ref{fig:latency:sensitivity} shows that application slowdowns on CXL significantly increase for devices at around 400\,ns of latency, based on data from prior work~\cite{liu2025cxl}.
Adapting applications~\cite{ahn2024examination,lee2023elastic}, OS tiering~\cite{tmts:asplos2023,vtmm:eurosys2023,vuppalapati2024tiered,xiang2024nomad,lee2023memtis,beyond-hotness}, or hardware tiering~\cite{zhong2024managing,hyperscale2023cxl} can only partially offset these slowdowns, especially at the tail.
Thus both datacenters and workloads look for low-latency access to CXL memory.

\subsection{Memory Capacity Pooling}
\label{sec:use-case-pooling}

Servers are typically provisioned with enough memory capacity to handle peak demands.
Unfortunately, capacity demand is highly variable across time and servers which leads to a large gap between average demand and peak~\cite{pond:asplos2023,coach-asplos}, causing low memory utilization.
Pooling capacity aims to \emph{multiplex these per-server peaks} so that capacity can be provisioned much closer to the average demand, thus provisioning less memory overall and improving memory utilization.
CXL enables memory pooling within a pod by allocating CXL-attached memory to servers (\eg at 1\,GiB granularity~\cite{pond:asplos2023}) and dynamically changing this allocation over time.

% It thus does not require any cross-host cache coherence.
% In CXL, dynamic allocation is implemented at a block size called \emph{extents}~\cite{das2024introduction,pond:asplos2023}.
% Importantly, memory is treated as a capacity resource, i.e., hosts do not typically care which extent is allocated to them.
% This enables significant flexibility in the pod memory allocation algorithm.
% % , which we further discuss in Section~\ref{sec:design}.

Prior work has shown significant cost savings from CXL capacity pooling and that these savings \emph{increase with the number of servers within a pod}.
% For example, we can reduce memory cost by multiplexing demands across cloud servers~\cite{pond:asplos2023, wahlgren2022evaluating}, lower cost and improve performance in in-memory databases~\cite{lee2023elastic,ahn2024examination}, lower cost for serverless function platforms~\cite{xu2024faasmem}, and improve application performance by adding more memory~\cite{gu2017efficient}.
% Figure~\ref{fig:poolingsavings} shows that the peak to average memory demand decays slowly when grouping servers in sets of increasing sizes.
% We need to group large numbers of servers to reduce outliers.
% This shows gains beyond prior work, in which savings increased from about 6--7\% to about 12\% of memory capacity when pooling between 4 to 16 servers~\cite{pond:asplos2023}.
Figure~\ref{fig:poolingsavings} shows that as pod size increases, aggregate peak memory demand within a pod approaches the aggregate average, with diminishing returns beyond about 96 servers. Thus, effectively reducing demand outliers requires pooling across dozens of servers.
% This shows gains beyond prior work, in which savings increased from about 6--7\% to about 12\% of memory capacity when pooling between 4 to 16 servers~\cite{pond:asplos2023}.

Beyond pod size, CXL access latency also affects pooling savings, since higher latency reduces the fraction of workloads that can tolerate CXL memory and thus limits how much memory can be pooled. Using workload sensitivity to CXL latency (Figure~\ref{fig:latency:sensitivity}), the measured latencies of \mpds and switches (Figure~\ref{fig:cxldev-with-lat}), and assuming a tolerable slowdown of 10\%, we estimate that 65\% of memory can be pooled and provisioned from \mpds, compared to 35\% when using switches.

%4 sockets: ~6–7% saved
%8 sockets: ~10–11%
%16 sockets: ~12–12.5%
%32 sockets: ~13%
%64 sockets: ~13.2–13.4%

\begin{comment}
\begin{figure}
    \centering
    \includegraphics[width=0.26\textwidth]{figures/poolingsavings.eps}
    \caption{Larger CXL pods lead to higher DRAM savings.}
    \label{fig:poolingsavings}
\end{figure}
\end{comment}

\subsection{Practical Shared Memory Use Cases}
\label{sec:use-case-sharing}

Besides memory pooling, where CXL-attached memory is allocated exclusively to individual servers, CXL today also allows multiple servers to concurrently access \emph{shared memory buffers} on CXL devices\footnote{CXL devices available today do \emph{not} support cross-server cache coherence. Although the CXL 3.0 specification introduces Back-Invalidate (BI), an optional cross-server hardware coherence mechanism~\cite{cxl-3.0}, it requires significant hardware changes on both processors and devices.}.
Shared CXL memory enables low-latency communication between servers via load/store instructions, and eliminates the need to serialize large and complex data structures when sharing them between servers, as required by Ethernet and RDMA.

% CXL today allows for shared memory among servers but does not support multi-server cache coherence\footnote{Although the CXL 3.0 specification introduces Back-Invalidate (BI), an optional cross-server hardware coherence mechanism~\cite{cxl-3.0}, it requires significant hardware changes on both processors and devices. Therefore, no CXL devices available today support BI.}.
% For this reason, CXL sharing is straightforward for read-only or single-writer scenarios.
% In particular, CXL is effective when used to overcome the need to serialize large and complex data structures, as required by standard networking, or when requiring very low latency.

\paragraph{Low-latency communication.}
% CXL can facilitate low-latency communication as we show in Section~\ref{sec:eval}.
Communication between two servers in a CXL pod can be implemented by having one server write to a message queue in a shared CXL buffer while the other busy-polls it~\cite{zhong2025oasis}.
In the best case, communication requires one CXL write and one CXL read, totaling roughly 600\,ns, which is orders of magnitude lower than typical RDMA latencies in datacenters (tens of \us~\cite{machnet}).

This is especially useful for distributed systems that rely on frequent small messages, including consensus protocols such as Viewstamped Replication~\cite{liskov2012viewstamped}, ZooKeeper~\cite{hunt2010zookeeper}, Raft~\cite{ongaro2014search}, proposer-acceptor messaging in Paxos~\cite{lamport1998parttime}, and HotStuff~\cite{yin2019hotstuff}.
% CXL-based distributed transaction systems similarly rely on a central coordinator, such as a transactional CXL controller~\cite{giles2024acid}.
More broadly, even complex communication patterns are commonly expressed using message-passing interfaces~\cite{clarke1994mpi,nvidia2016nvidia,cowan2023mscclang,shah2023taccl}.

Many high-availability scale-out systems operate at \emph{modest cluster sizes, typically between 3 and 16 servers}:
MySQL InnoDB Cluster (3--7 nodes)~\cite{mysql_innodb_cluster},
MongoDB replica sets (3--7)~\cite{mongodb_replica_members},
SAP~HANA TDI (up to 16)~\cite{sap_kba_3557729,sap_tdi_faq_pdf},
Amazon Aurora (up to 15 read replicas)~\cite{aws_aurora_replicas},
Redis Cluster (6 nodes)~\cite{redis_cluster_scale},
Bigtable (3+ nodes)~\cite{gcp_bigtable_production_min_nodes},
and primary-backup replication~\cite{author2025thesis,cohen2024dictionary,zhu2024lupin,cxl-shm,MemVerge_GISMO_ISC2023,Tian_GISMO_FMS2023}.

% Larger scale-out systems are often driven by data analytics workloads. Examples include Azure SQL Database Hyperscale, which supports up to 30 named read replicas per primary~\cite{azure_hyperscale_named_replicas}, Postgres sharding with Citus where dozens of workers are common~\cite{citus_cluster_mgmt}, cloud data warehouses such as Amazon Redshift, which support up to 128 servers per cluster~\cite{redshift_limits_ra3}, and SAP~HANA analytic scale-out deployments with up to 94 servers~\cite{lenovo_hana_solution_brief}. Even larger systems can span hundreds to thousands of hosts, but these typically scale by adding additional clusters rather than expanding a single cluster.

%Pair-wise communication is also useful for multi-socket CXL systems like StarNUMA~\cite{cho2024starnuma}.

%A related use case is migration of virtual machines, containers, or processes across cloud hosts~\cite{vmmigration}.
%Fail-over requires at least one other host to have access to the same memory as the failing host.

\paragraph{Eliminating serialization.}
Sharing large and complex data structures over Ethernet and RDMA requires expensive serialization and compression. Prior work shows that these overheads account for 6--10\% of CPU time in RPC-heavy services and up to 20\% for small 1\,KB payloads~\cite{sriraman2020accelerometer,kalia2019erpc,raghavan2021breakfast,huye2023meta,seemakhupt2023cloudscale,pourhabibi2021cerebros,ma2024hydrarpc,rpcool,zhang2024dmrpc}. In CXL pods, \emph{these serialization and compression overheads can be avoided} by allocating pairwise shared CXL buffers between servers~\cite{ma2024hydrarpc,rpcool,zhang2024dmrpc,baumstark2024so,cxl-shm}.

\begin{table}[t]
\small
\centering
\begin{tabular}{ll}
\hline
\textbf{Parameter} & \textbf{Description} \\
\hline
$X$ & Number of CXL ports per server \\
$N$ & Number of CXL ports per \mpd \\
$S$ & Number of servers in a pod \\
$M$ & Number of \mpds in a pod \\
$X_i$ & Number of server ports to island-specific \mpds  \\
% $e_k$ & Expansion at scale $k$ \\
% $D_k$ & Max aggregate memory demand of any $k$ servers \\
\hline
\end{tabular}
\vspace*{-0.5em}
\caption{\mpd topology notation.}
\label{tab:topo-parameters}
\vspace*{-1.5em}
\end{table}

\begin{table}[t]
\centering
\footnotesize
\setlength{\tabcolsep}{4pt}
\renewcommand{\arraystretch}{1.1}

% Dark colors
\definecolor{darkred}{RGB}{190,60,60}
\definecolor{darkgreen}{RGB}{40,140,40}
\definecolor{darkyellow}{RGB}{215,140,0}

\begin{tabular}{lcc}
\toprule
\textbf{\mpd Topology} & \textbf{Pooling} & \textbf{Communication Latency} \\
\midrule
Fully-connected~\cite{pond:asplos2023} ($S{=}4$)
  & \textcolor{darkred}{Poor}
  & \textcolor{darkyellow}{Low (4)} \\

BIBD~\cite{pignolet-bibd,cypher2015configuring} ($S{=}25$)
  & \textcolor{darkred}{Poor}
  & \textcolor{darkgreen}{Low (25)} \\

Expander~\cite{singla2012jellyfish,valadarsky2016xpander} ($S{=}96$)
  & \textcolor{darkgreen}{Optimal}
  & \textcolor{darkred}{High} \\

\textbf{\sysname} \textbf{($S{=}96$)}
  & \textbf{\textcolor{darkgreen}{Near Optimal}}
  & \textbf{\textcolor{darkgreen}{Low (16)}} \\
\bottomrule
\end{tabular}

\vspace*{-0.5em}
\caption{Memory pooling effectiveness and communication latency of \mpd topologies with $N=4$ and $X\le8$ constraints. ``Low ($k$)'' denotes low-latency communication among $k$ servers, while ``High'' indicates multi-hop server-level forwarding.}
\label{tab:topo-tradeoff}
\vspace*{-1em}
\end{table}

\section{\sysname Design}\label{sec:design}

Rather than relying on CXL switches, \sysname builds on \mpds as a low latency and cost-effective building block for CXL pods.
We seek to fulfill the following three goals:
\begin{denseenum}
\item \textbf{Effective memory pooling}, which requires a large number of interconnected servers within a CXL pod~(\S\ref{sec:use-case-pooling}).
\item \textbf{Low-latency communication}, which requires servers to directly share an \mpd~(\S\ref{sec:use-case-sharing}).
\item \textbf{Practical deployability}, which limits us to short copper CXL cables ($\le1.5$\,m), \mpds with $N=4$ ports, and at most eight CXL links per server~(\S\ref{sec:bg}).
\end{denseenum}

% Prior \mpd-based CXL pod designs assume fully-connected topologies, where every \mpd connects to every server~\cite{pond:asplos2023}, limiting pod size to the \mpd port count and failing to achieve effective memory pooling (goal \#1). To overcome this limitation, 

\sysname uses \emph{sparse \mpd topologies}, where each \mpd connects to only a subset of servers, enabling larger pods despite limited \mpd port counts.
However, designing sparse \mpd topologies that simultaneously satisfy goals \#1 and \#2 is challenging because they favor conflicting properties.
% Low-latency communication (goal~\#2) requires pairs of servers to \emph{collide} on a common \mpd, while effective pooling (goal~\#1) prefers avoiding \mpd collisions among sets of "hot" servers so that their excess memory demand can be spread across more \mpds.

We first describe the conflicting topology requirements for effective memory pooling and low-latency communication~(\S\ref{sec:tension}).
We then present \sysname's logical server-to-\mpd topologies~(\S\ref{sec:logicaltopo}), physical layout in racks~(\S\ref{sec:physicallayout}), and software stack~(\S\ref{sec:sw}).

% We describe the tension between these design goals and then move from graph representations to physical realizations and the software stack.

%Non goal: global sharing between all hosts

\subsection{Tension \& Balancing Use Cases}
\label{sec:tension}

% Central to \sysname's feasibility is the assumption that every host provides $X>1$ independent CXL ports, allowing each host to connect simultaneously to multiple \mpds.
% Current hardware commonly offers $X=8$ ports per host (Section~\ref{sec:bg}), though we demonstrate that $X=4$ ports per host is sufficient to realize \sysname's potentials (Section~\ref{sec:eval}).
% Multiple host ports enable each host to connect to multiple \mpds which in turn connect to different subsets of hosts.
% For simplicity and clarity, we assume single-socket servers\footnote{\sysname topologies also apply to multi-socket servers. In this case, we still seek to connect pairs of CPU sockets with each other. \yuhong{Why do we want to connect the two sockets of a host via a MPD?} However, we incorporate the cache-coherent processor interconnection network to reduce required CXL links.} where each CPU constitutes a distinct host.

To show the tension between memory pooling (goal~\#1) and low-latency communication (goal~\#2), we model a CXL pod as a bipartite graph with two vertex sets: servers and \mpds.
Edges represent CXL links between servers and \mpds. Each server has degree $X$, corresponding to the number of CXL ports per server, and each \mpd has degree $N$, corresponding to the number of ports per \mpd. The notation is summarized in Table~\ref{tab:topo-parameters}.
For example, a fully-connected topology is a complete bipartite graph where the \mpd port count ($N$) equals the number of servers ($S$).
In contrast, in a sparse topology, $S$ can be much larger than $N$.
% For simplicity, we assume single-socket servers~(\S\ref{sec:discussion}).
% \footnote{\sysname topologies also apply to multi-socket servers. In this case, we still seek to connect pairs of CPU sockets with each other.
% % \yuhong{Why do we want to connect the two sockets of a server via a \mpd?}
% However, we incorporate the cache-coherent processor interconnection network to reduce required CXL links.} where each CPU constitutes a server.

The tension between memory pooling and low-latency communication is because they have conflicting requirements in terms of \emph{\mpd overlap}, \ie whether servers connect to a common \mpd.
We will show why they favor opposing overlap properties as well as why existing topologies (\eg datacenter network topologies~\cite{singla2012jellyfish,valadarsky2016xpander,bcube,pignolet-bibd,cypher2015configuring,copyset}) fail to balance the tension.

% The tension between memory pooling and low-latency communication is because they have conflicting requirements in terms of \emph{\mpd collisions}, \ie whether servers connect to a common \mpd.
% % Low-latency communication favors pairwise \mpd collisions between any two servers, whereas memory pooling disfavors \mpd collisions among "hot" servers so that their excess memory demand can be spread across as many \mpds as possible.
% % \yuhong{briefly describe the conflicting collision requirement, and we use island to do the tradeoff.}
% We will show why they prefer conflicting collision properties as well as why existing topologies (\eg datacenter network topologies~\cite{singla2012jellyfish,valadarsky2016xpander,bcube,pignolet-bibd,cypher2015configuring,copyset}) fail to balance the conflicts.

% We seek to create a topology that works for memory pooling as well as data sharing and low-latency communication use cases.
% A CXL pod topology can generally be represented as a graph where edges connect hosts to \mpds.
% Since there are no host to host edges, any two hosts are at least two hops away, traversing an \mpd.
% The same is true for \mpd to \mpd connections.

\subsubsection{Communication Favors Pairwise Overlap}
\label{sec:design-comm-bibd}
Low-latency communication over CXL requires pairs of servers to be able to issue load/store instructions to the same \mpd directly.
Although servers without a common \mpd can still communicate through intermediate hops by forwarding messages through other servers and \mpds, we find that this server-level forwarding loses CXL's latency advantages over RDMA~(\S\ref{sec:eval-hw}).

% CXL is uniquely suited for serialization-free data sharing and low-latency small-request communication.
% This requires pairs of hosts to be able to issue loads/stores to the same MPD directly.
% We show in Section~\ref{sec:eval} that any forwarding through other hosts loses CXL's unique advantages.
% Even forwarding through switches can significantly inflate latency (\S\ref{sec:eval}).

Therefore, the ideal graph property for low-latency communication is \emph{pairwise \mpd overlap}, where every pair of servers in a pod connects to at least one common \mpd.

The best graphs for this property are Balanced Incomplete Block Design (BIBD) graphs~\cite{fisher_yates_1938,bose_1939,shrikhande_1973}, which have also been used for datacenter networks~\cite{pignolet-bibd,cypher2015configuring}.
BIBD is a classical combinatorial design used to construct bipartite graphs in which every pair of vertices on one side (servers) connects to exactly one common vertex on the other side (\mpds)\footnote{BIBD uses the parameter $\lambda$ to control the number of vertices on one side of the bipartite graph shared by each pair of vertices on the other side. We set $\lambda=1$ to enforce the pairwise overlap requirement.}.

With $N=4$-port \mpds and using $X\le8$ ports per server, BIBD yields three pod topologies with different sizes: 13 servers ($X=4$), 16 servers ($X=5$), and 25 servers ($X=8$).
Note that the 25-server graph is the largest BIBD graph with $N=4$ and $X\le8$, implying that larger pods cannot satisfy the pairwise overlap property.
As we show later (Table~\ref{tab:topo-tradeoff}), although the 25-server BIBD graph maximizes the number of servers that can communicate at low latency (goal~\#2) under practical CXL constraints (goal~\#3), it performs poorly for memory pooling (goal~\#1).

\subsubsection{Pooling Disfavors Overlap Among Hot Servers}

Memory pooling reduces overall memory capacity by multiplexing server demand peaks~(\S\ref{sec:use-case-pooling}).
Pooling savings are determined by the peak memory usage across \mpds because \mpds must be provisioned to handle the worst-case load.
Therefore, our goal is to minimize the \emph{peak memory usage across \mpds} in a pod.

% Pooling reduces overall memory capacity by multiplexing server demand peaks \cite{pond:asplos2023}.
% With \(\frac{1}{3}\)–\(\frac{2}{3}\) of memory capacity on CXL (Section~\ref{sec:usecases}), our goal is to reduce the \emph{peak memory usage across \mpds}.

% Our main insight is that pooling \emph{does not require a single monolithic pool} accessed by all hosts.
% % The key is to think about hosts allocating CXL capacity from \mpds over time, not as a static one-time assignment.
% The key is that hosts can observe the load on their connected \mpds and allocate from the least-loaded \mpds, and thus can balance load across \mpds.
% Thus, what matters is that when some servers run “hot,” their excess demand can spill into \emph{enough distinct} \mpds so that no small subset of \mpds becomes the bottleneck.
% Figure~\ref{fig:mem-concentration} shows that the top 11\% of servers with the highest memory demand account for 50\% of the total memory demand across all servers at the peak in a cluster at \azure.

% \begin{figure}[!t]
%     \centering
%     \includegraphics[width=0.46\textwidth]{figures/memory-concentration.pdf}
%     \vspace*{-0.5em}
%     \caption{Cumulative contribution of peak memory demand as a function of the percentage of hot servers, showing that a small fraction of servers accounts for a disproportionate share of total memory usage. \yuhong{Should we keep this figure?}}
%     \label{fig:mem-concentration}
% \end{figure}

Memory demand in datacenters is highly spiky across servers~\cite{coach-asplos,hadary2020protean,pond:asplos2023}.
As a result, peak \mpd usage is often dominated by a small subset of ``hot'' servers with high memory demand in a pod.
To maximize pooling savings, the topology should therefore \emph{avoid \mpd overlap among worst-case hot server sets}, allowing their excess demand to be spread across as many distinct \mpds as possible, since demand patterns are not known in advance.

\mpd overlap among servers can be characterized using graph \emph{expansion}~\cite{hoory2006expander}.
For a given subset size $k$, we define the graph's expansion $e_k$ as the minimum number of distinct \mpds connected to any subset of $k$ servers~\cite{valadarsky2016xpander,hoory2006expander}.
Expansion directly yields a lower bound on peak \mpd usage\footnote{The theorem and its proof appear in Appendix~\ref{app:expansion}.}.
\emph{Expander} graphs, including random graphs (\eg Jellyfish~\cite{singla2012jellyfish}) and Ramanujan graphs (\eg Xpander~\cite{valadarsky2016xpander}), provide asymptotically optimal expansion for fixed $X$ and $N$~\cite{hoory2006expander}. Moreover, larger expander graphs achieve better expansion than smaller ones.
Under practical cabling constraints (goal~\#3), expander graphs with up to 96 servers are feasible~(\S\ref{sec:physicallayout}).

\begin{figure}[!t]
    \centering
    \includegraphics[width=0.8\columnwidth]{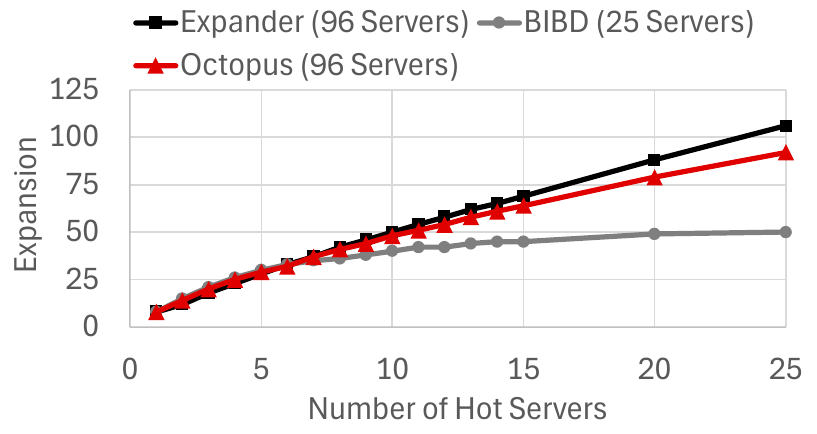}
    \vspace*{-0.8em}
    \caption{Expansion across topologies as the number of hot servers varies. Expansion denotes the number of distinct \mpds connected to the worst-case hot server set.}
    \label{fig:hot-reachability}
    \vspace*{-0.8em}
\end{figure}

Figure~\ref{fig:hot-reachability} shows that a 96-server expander graph achieves stronger expansion than the 25-server BIBD graph. While expander graphs are well suited for memory pooling, they do not satisfy the pairwise overlap property required for low-latency communication. In the 96-server expander graph, worst-case communication between two servers requires forwarding through two intermediate servers~(Table~\ref{tab:topo-tradeoff}).

\begin{comment}

For simplicity, we primarily consider the minimum number of distinct \mpds reachable from any server.
This is a natural proxy for the worst-case hot server set, since a server with poor reachability can easily be part of a hot set.
We visualize reachability for a 16-host topology in Figure~\ref{fig:acadia16}; this topology is sub-optimal as server ports are wasted on redundant paths between any host and its reachable \mpds.
Figure~\ref{fig:reachability} shows how larger 96-host topologies with specific structures (discussed below) significantly increase reachability.
\yuhong{Replace this with the new reachability experiment.}

In general, pooling improves with reachability and thus favors three key graph properties.
First, larger topologies always offer more distinct \mpds overall.
Second, low-diameter topologies ensure that no server is “far away” from many \mpds.
Third, high symmetry ensures that all servers have similar reachability.

\end{comment}

\begin{figure}[!t]
    \centering
    \includegraphics[width=0.41\textwidth]{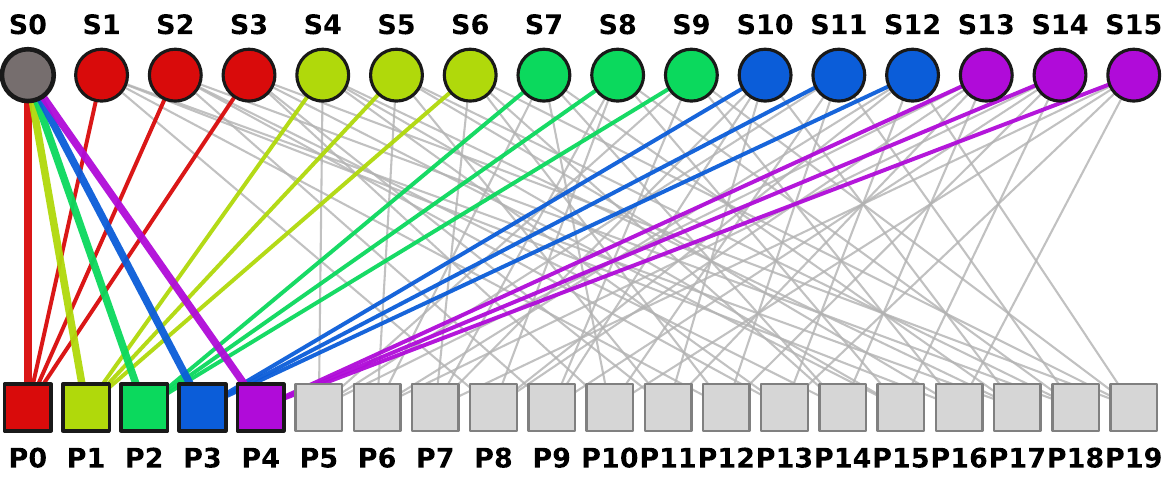}
    \vspace*{-0.5em}
    \caption{An \sysname island guarantees pairwise \mpd overlap, \ie every pair of servers connects to exactly one common \mpd. The \mpds and links connecting server ``S0'' to other servers are highlighted; the same applies to each server.
    % For pooling, we largely care about the number of distinct \mpds reachable from any host (here, H0).
    % The topology shows that any MPD can be reached in four hops, but there are many redundant paths which shows that this is suboptimal topology for \mpd reachability and thus pooling.
    }
    \vspace*{-1em}
    \label{fig:acadia16}
\end{figure}

% \begin{figure}[h]
%     \centering
%     \includegraphics[width=0.46\textwidth]{figures/reachability_bibd.pdf}
%     \caption{\mpd reachability defines how well a memory demand hot spot can be dissipated across \mpds.
%     A diameter 2 topology with \(X=8\) and \(N=4\) can only reach 50 \mpds within 3 edges from any host, while larger topologies can reach almost 100 \mpds at the same distance and continue to grow.}
%     \label{fig:reachability}
% \end{figure}

\subsection{Logical Network Topology}
\label{sec:logicaltopo}

To balance the tension between memory pooling and low-latency communication, we observe that low-latency communication operates at smaller scales (\eg 16 servers~(\S\ref{sec:use-case-sharing})) than effective memory pooling, which begins to show meaningful gains at around 64 servers~(\S\ref{sec:eval-sim}).

\sysname leverages this observation by organizing each pod into multiple \emph{islands} of highly connected servers that provide low-latency intra-island communication, while maintaining sufficient inter-island connectivity to increase expansion and enable effective memory pooling~(Figure~\ref{fig:interconnect-island}).
% \yuhong{shoud visualize.}

Each island satisfies the pairwise \mpd overlap property.
% and is intentionally kept small (\eg 16 servers) to limit collisions among hot servers.
A subset of $X_i$ server ports on each server is dedicated to connecting to \emph{island-specific \mpds}, while the remaining $X - X_i$ ports connect to additional \mpds that interconnect islands and provide the expansion required for pooling.

One might expect \sysname's hierarchical design to reduce expansion. However, Figure~\ref{fig:hot-reachability} shows that a 96-server \sysname topology achieves expansion close to that of a 96-server expander graph.
\sysname trades off the size of the low-latency communication domain to achieve this expansion.
In particular, using $X_i=8$ yields an island of 25 servers, while we empirically find that $X_i=5$ provides strong expansion with islands of 16 servers, a 36\% reduction in low-latency communication domain. 
This tradeoff is reasonable given the typical cluster sizes for low-latency communication~(\S\ref{sec:usecases}).

% \sysname topologies form a family of designs based on the number of islands.
% Assuming $X_i=5$, we recommend pods of 25 (single island), 64 (four islands), 96 (six islands), and 128 (eight islands) hosts.
% We next discuss in more detail how to construct and interconnect islands.

\begin{table}[t]
\centering
\footnotesize
\setlength{\tabcolsep}{6pt}
\renewcommand{\arraystretch}{1.15}
\begin{tabular}{c c c c}
\toprule
\# islands & \# servers per island & Server count ($S$) & \mpd count ($M$) \\
\midrule
1 & 25 & 25 & 50 \\
\rowcolor{gray!10}
4 & 16 & 64 & 128 \\
\textbf{6} & \textbf{16} & \textbf{96} & \textbf{192} \\
\bottomrule
\end{tabular}
\vspace*{-0.5em}
\caption{\sysname defines \mpd pod designs parameterized by island count. The default pod (shown in \textbf{bold}) has 6 islands and 96 servers. All designs use $X{=}8$ ports per server and $N{=}4$-port \mpds.}
\vspace*{-0.5em}
\label{tab:octopus-topo}
\end{table}

\begin{figure}
    \centering
    \includegraphics[clip,width=0.9\columnwidth]{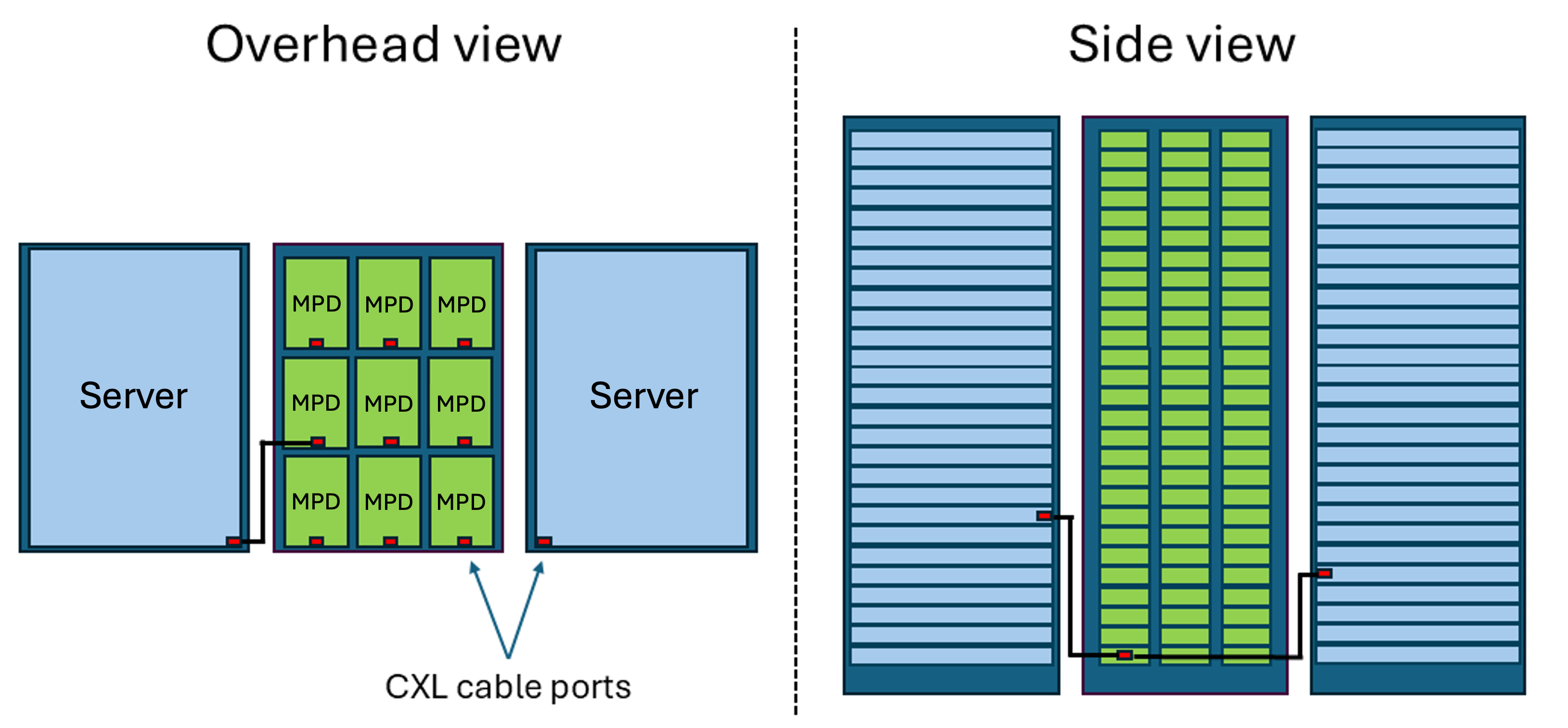}
    \vspace*{-0.5em}
    \caption{A 3-rack \sysname configuration.}
    \label{fig:rack_config}
    \vspace*{-1em}
\end{figure}

\sysname forms a family of \mpd pod designs parameterized by the number of islands~(Table~\ref{tab:octopus-topo}).
\sysname includes a single-island configuration with 25 servers, where all server ports are used for intra-island connectivity.
For multi-island pods, we allocate $X_i=5$ ports per server for intra-island connectivity, yielding feasible pod sizes of 64 (four islands) and 96 (six islands).
We use the 96-server pod by default, as it achieves near-optimal memory pooling (goal~\#1) and low-latency communication within each island (goal~\#2) while satisfying deployability constraints (goal~\#3).

We next describe how islands are constructed and interconnected in more detail.

\subsubsection{Intra-Island Connectivity}

Each island is a subgraph in which every pair of servers connects to exactly one common \mpd~(\S\ref{sec:design-comm-bibd}), enabling low-latency communication within the island without intermediate hops~(Figure~\ref{fig:acadia16}).
In multi-island \sysname pods, each island is a BIBD graph with 16 servers and 20 island-specific \mpds, where each server uses $X_i=5$ ports to connect to the island-specific \mpds.
The larger 25-server BIBD graph~(\S\ref{sec:design-comm-bibd}) is used only in the single-island \sysname configuration, as it consumes all eight CXL ports per server for island-specific \mpds, leaving no ports available for inter-island connectivity.

\subsubsection{Inter-Island Connectivity}

% The main purpose of inter-island connectivity is to increase the expansion of hot server sets for pooling.
% We use dedicated "external" \mpds and the remaining $X - X_i$ host ports.
% For $X_i=5$, every island has 48 external links.
% For the 96-server \sysname pod with six islands, this means that there are 72 external \mpds, \ie 37.5\% of all \mpds are external.

The main purpose of inter-island connectivity is to increase the expansion of hot server sets for memory pooling.
We achieve this using dedicated "external" \mpds and the remaining $X - X_i$ server ports. With $X_i=5$, each island has 48 external links.
In the 96-server \sysname pod with six islands, this corresponds to 72 external \mpds, or 37.5\% of all \mpds.

% We find that inter-island connectivity is a challenging combinatorial design problem.
% An effective starter heuristic is that every external \mpd should connect to servers from \emph{different} islands.
% However, we find that we must actively enforce high symmetry.
% Specifically, each pair of islands must intersect in the same number of \mpds.
% Finally, any two servers from different islands should share at most one external \mpd to cap worst-case \mpd collisions.
% In short, we require uniformity at multiple scales: across islands, across island pairs, and across server pairs.

Designing inter-island connectivity is a challenging combinatorial problem.
A natural starting heuristic is to connect each external \mpd to servers from \emph{different} islands.
However, achieving good expansion requires enforcing strong symmetry.
Specifically, each pair of islands must share the same number of external \mpds, and any two servers from different islands should share at most one external \mpd to bound worst-case \mpd overlap.
In short, the design must maintain uniformity across islands, island pairs, and server pairs.

% We address this design problem using a two-level approach.
% At the first level, we find a block design (similar to BIBD) for each external \mpd that defines which islands should interconnect.
% In practice, the block design cannot always be solved exactly, so we fall back to a round-robin heuristic.
% The second level assigns servers from the sampled island to the specific \mpd ports.
% With $X_i=5$, there are three ports per host.
% We proceed in three rounds, where in each round every host is used exactly once.
% We additionally ensure that any two servers from different islands share at most one external \mpd.

We address this design problem using a two-level approach.
At the first level, we select the set of islands connected by each external \mpd using a block design similar to BIBD. When an exact solution is not feasible, we fall back to a round-robin heuristic.
At the second level, we assign servers from the selected islands to the specific ports of each \mpd.
With $X_i=5$, each server has three external ports, and we perform the assignment in three rounds such that each server is used exactly once per round.
We additionally enforce that any two servers from different islands share at most one external \mpd.

%Optimality fior reachability would mean that each island should minimize how many wasted edges it has, perhaps less symmetry

\subsection{Physical Layout}\label{sec:physicallayout}

The physical layout of an \sysname pod is largely driven by CXL cable lengths, cable routing, and constraints of typical datacenter rack layouts.
Reasonable cable lengths should be less than 1.5\,m~(\S\ref{sec:bg}).
To deploy pods with up to 96 servers in a datacenter row, we propose a 3-rack pod designs where \mpds are places in the middle rack and servers placed in adjacent racks.
Figure~\ref{fig:rack_config} sketches this design.
The middle rack can hold a varying number of \mpds in each slot, depending on the \mpd size (\eg $N=4$ vs $N=8$).
The dimension of each standard rack slot is approximately 100 $\times$ 60 $\times$ 5\,cm.

To further optimize cable lengths, we assume servers where CXL edge connectors are located at the front corner of the server chassis closest to the \mpd rack, directly on the motherboard.
This is a similar requirement to those imposed by the OCP NIC 3.0 specification~\cite{ocp_nic_3_0} which also requires PCIe5 ports at the front of the server.
Based on existing designs, we assume that CXL ports on the \mpd side are in the front-middle of each \mpd.

We validate achievable \sysname pod designs in \S\ref{sec:eval-phy}.

% Daniel editing SW stack just now
\subsection{\sysname Software Stack}
\label{sec:sw}

\begin{figure}
  \centering
  \begin{subfigure}{0.22\textwidth}
    \centering
    \includegraphics[width=0.99\textwidth]{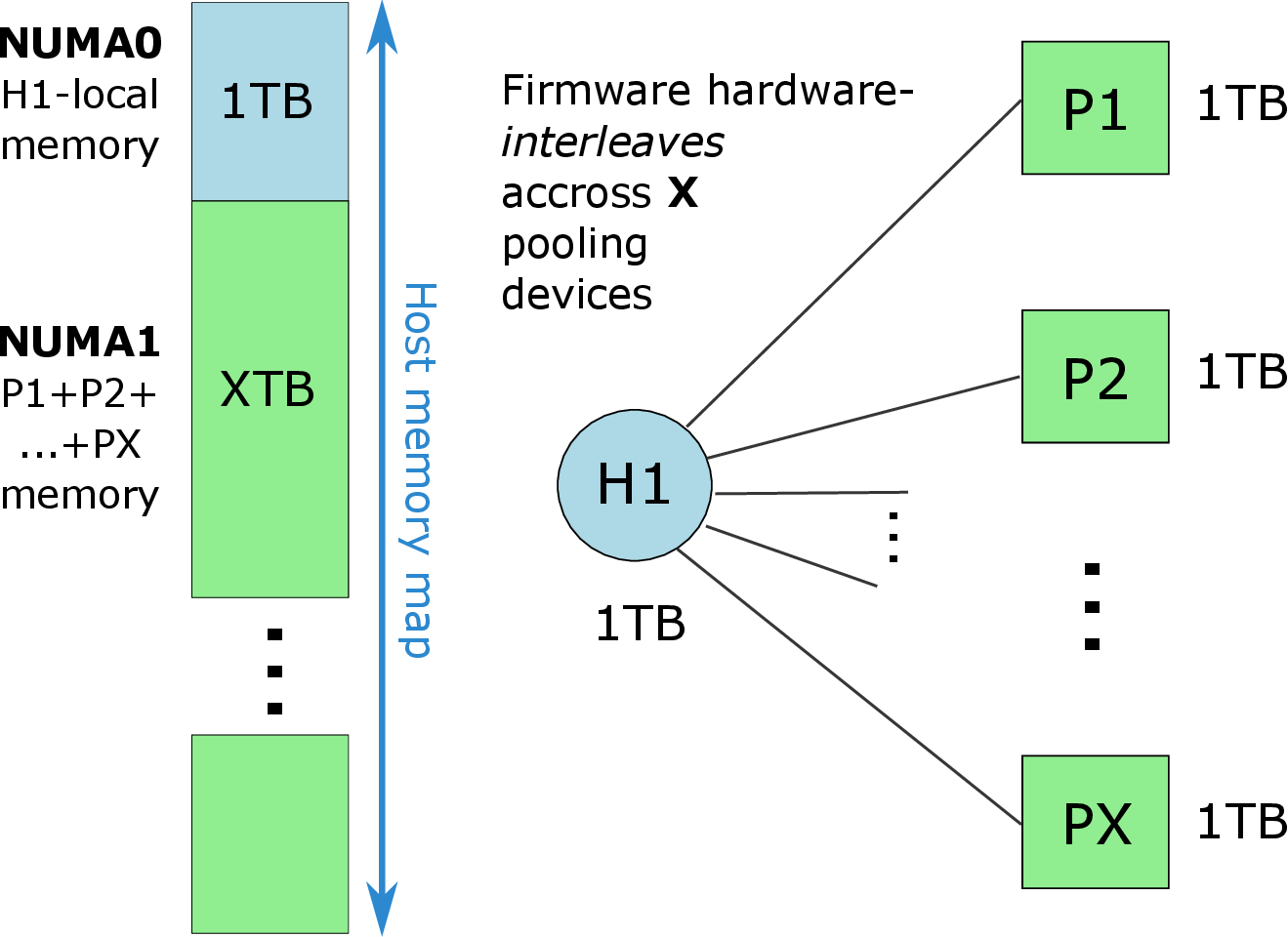}
    \caption{Fully-connected.}
    \label{fig:fcsw}
  \end{subfigure}
  \begin{subfigure}{0.22\textwidth}
    \centering
    \includegraphics[width=0.99\textwidth]{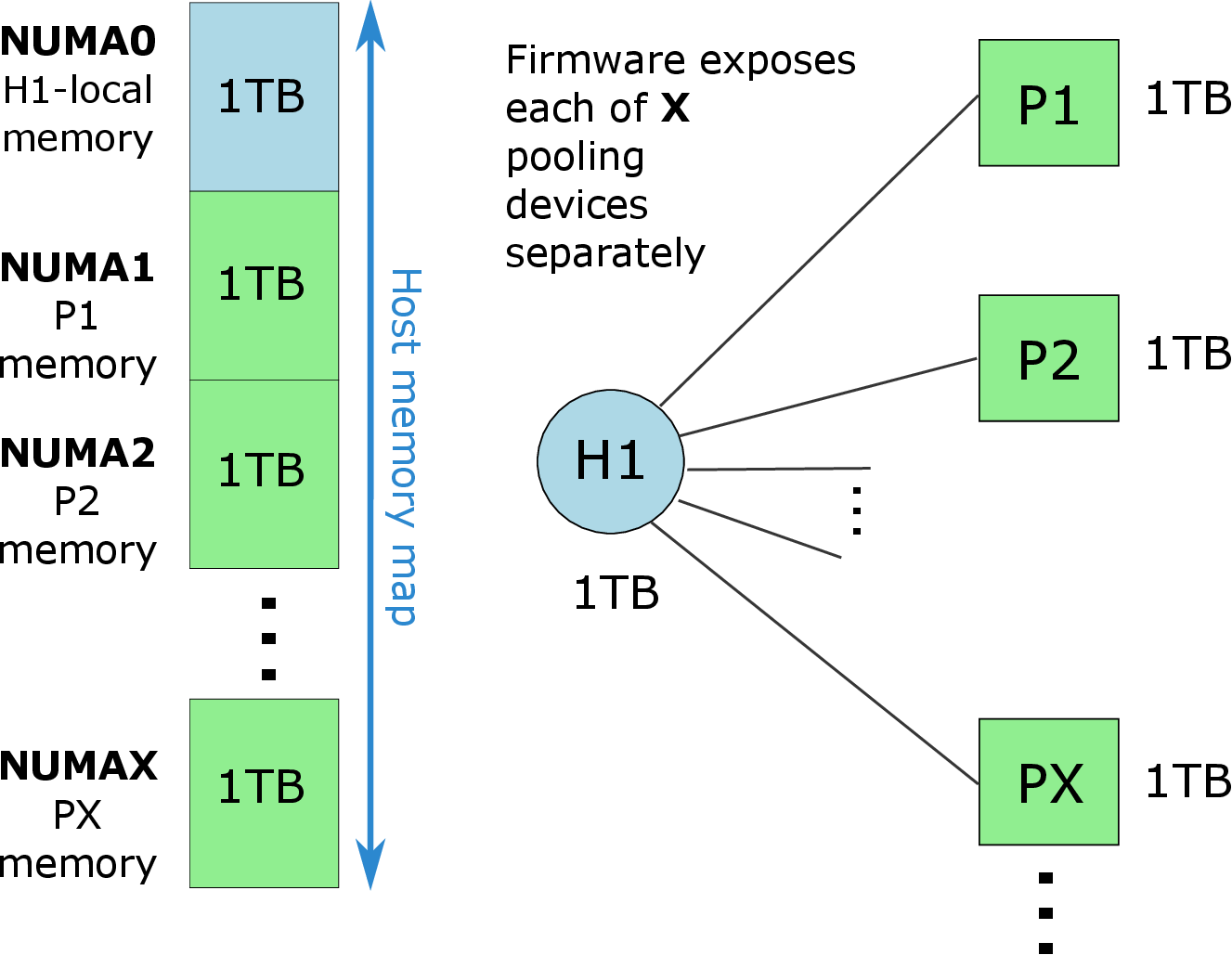}
    \caption{\sysname (sparse).}
    \label{fig:sysfw}
  \end{subfigure}
  \vspace*{-0.5em}
  \caption{Host view of CXL memory under interleaving vs. \sysname.}
  \label{fig:hostsw}
  \vspace*{-0.5em}
\end{figure}

\textbf{API and exposure.}
In fully-connected \mpd pods, server firmware interleaves \mpds at 256\,B granularity~\cite{CXL2spec}, yielding one large pool (Figure~\ref{fig:fcsw}). 
\sysname disables this interleaving in firmware and exposes each CXL port as a distinct NUMA node (Figure~\ref{fig:sysfw}).
This lets software target a specific \mpd for capacity balancing and for sharing with peer servers that connect to the same \mpd.
A datacenter control plane (\eg Borg/Protean-like) assigns server IDs and disseminates the \mpd pod topology and each server's \mpd set~\cite{hadary2020protean,tirmazi2020borg}.

\paragraph{Pooling policy.}

When allocating CXL memory, each server allocates from the \emph{least-loaded} connected \mpd, prioritizing those with the most available capacity. This reduces allocation failures caused by individual \mpds becoming fully utilized, without requiring global defragmentation.

% \paragraph{Bandwidth scaling.}
% When a single $\times 8$ link is insufficient, \sysname interleaves memory at page granularity (4\,KB or 2\,MB) across multiple \mpds or applies bandwidth-aware tiering~\cite{vuppalapati2024tiered}. We prefer using the fewest \mpds that satisfy the observed bandwidth demand to facilitate capacity balancing.

% \paragraph{Sharing within an island.}
% Any two hosts in an island share at least one \mpd. \sysname places per-\mpd input queues there and uses a lightweight poll loop per queue. RPCs use these queues; shuffles and similar collectives write to all targets’ queues, then read back. No server forwarding or multi-hop paths are required.

\section{Evaluation}\label{sec:eval}

\begin{comment}
\begin{table}[t]
  \centering
  \caption{RPC RTT percentiles across paths.}
  \label{tab:rpc-rtt}
  \begin{tabular}{l
                  S[table-format=2.2]
                  S[table-format=2.2]
                  S[table-format=2.2]
                  S[table-format=2.2]}
    \toprule
    \textbf{Path} & \textbf{P0 [\si{\micro\second}]} & \textbf{P50 [\si{\micro\second}]} & \textbf{P90 [\si{\micro\second}]} & \textbf{P99 [\si{\micro\second}]} \\
    \midrule
    Direct MPD         & 0.9  & 1.2  & 1.4  & 1.7  \\
    1-host forward     & 2.8  & 3.8  & 4.3  & 4.7  \\
    2-host forward     & 4.4  & 5.4  & 6.0  & 6.4  \\
    3-host forward     & 6.2  & 7.7  & 8.4  & 11.2 \\
    \addlinespace
    RDMA (idle)        & 3.8  & 3.81 & 3.82 & 3.83 \\
    CXL switch (lab)   & 2.3  & 2.9  & 3.2  & 3.5  \\
    CXL switch (theory)\hspace*{-3em}& 1.7  & 2.2  & 2.6  & 3.1  \\
    \bottomrule
  \end{tabular}
\end{table}
\end{comment}

Our evaluation of \sysname includes a small-scale hardware prototype evaluation~(\S\ref{sec:eval-hw}), scaled-up simulations~(\S\ref{sec:eval-sim}), physical layout validation~(\S\ref{sec:eval-phy}), and cost modeling~(\S\ref{sec:cost-cmp}).

%What are the implications of our design for memory-sharing use cases such as pair-wise communication, shuffle, and broadcast?
%\emph{\sysname works well for pair-wise communication and shuffle, but has significantly longer broadcast completion times than \baseline topologies.}

\subsection{Evaluation Setup}
\label{sec:eval-setup}

\paragraph{Hardware.}
We have access to three pre-production \mpds with two $\times 8$ CXL 2.0 ports (\ie $N=2$).
To connect them, we soldered custom risers to route CXL control signals to all ports.
We use three servers each with an AMD EPYC 9825 Processor, local DDR5-5600 memory, and a 100\,Gbit Mellanox CX5 NIC.
Each server has 64 CXL lanes, but we connect only 16 lanes (two $\times 8$ lanes) per CPU to form a simplified \sysname island ($X=N=2$): each pair of servers shares one \mpd.
All servers additionally connect with a 100\,Gbit Arista 7060X series switch and run an updated Ubuntu 24.04.

\paragraph{RPC.} We implement a simple CXL-based RPC by passing a message that contains the RPC parameters from one server to another and returning a new message containing the return value back to the initiating server.
To pass a message, the sender first writes the message to a \mpd, while the receiver busy polls the \mpd to retrieve the message.
The message can be a copy of parameters/return values or a pointer to them.
To measure the RPC latency with RDMA, we use the \texttt{send} RDMA verb to send RPC parameters and to send return values between two servers.
We also measure RPC latency with a user-space networking stack~\cite{fried2024making}.

%\textbf{RPCs}
%For simplicity, we assume that RPCs requests have to reach a core, and that the core immediately sends its response, without computation.
%We implement RDMA RPCs using the ib\_send\_lat tool~\cite{perftest}.

\paragraph{Simulations.}
Our simulations study \mpd topologies with between 2 to 8 \mpd ports ($N$), though we focus on $N=4$.
We consider server ports ($X$) between 4 to 16, though we focus on $X=8$.
Our main configuration is the 96-server \sysname pod (\sysname-96) consisting of six islands of 16 servers each.

Memory pooling simulations play back VM memory demand traces from \azure~\cite{coach-asplos}.
% \footnote{\azure is a large cloud provider. Name anonymized for submission.}
We collected trace data over a two-week period in December 2024 from thousands of cloud VMs.
For each VM, we record its allocation and deallocation times, resource allocation, and the hosting server.

When a VM launches, it greedily allocates memory from the least utilized connected \mpds~(\S\ref{sec:sw}), potentially spanning multiple \mpds, and releases memory on termination. We randomly group servers into CXL pods, replay the traces, and record the peak usage across all \mpds. This peak determines per-\mpd capacity, with lower peaks indicating better pooling.

\paragraph{Physical layout model.}
We validate the physical realization of \sysname topologies within the 3-rack layout~(\S\ref{sec:design}) under CXL cable length constraints by modeling server and \mpd placement as a satisfiability (SAT) problem.
Each potential server and \mpd location within the racks is modeled as a 3-D coordinate that specifies the location of the ports for each server/\mpd slot.
%For large topologies where the host count exceeds the available rack slots, we model two hosts sharing a rack slot, with half-slot host ports located at the edge of the slot from that is closest to the \mpd rack.

The cable length limit is modeled as a three-dimensional Manhattan distance between server and \mpd ports, and is enforced through constraints on the one-to-one mapping of logical \mpd and server pairs
% -- as defined by a given logical topology -- 
to physical locations in the racks.
% Specifically, if logical \mpd $i$ and server $j$ are connected in the topology, any physical pair of \mpd $\alpha$ and server $\beta$ that have a Manhattan distance greater than the cable length must not both be mapped to $i$ and $j$, \ie $\neg (i \rightarrow \alpha) \lor \neg (j \rightarrow \beta)$.
If a mapping that satisfies the constraints is found, then the mapping should be realizable physically, modulo cabling complexity. 
We implement the model in PySAT~\cite{itk-sat24} and use MiniSat 2.2 to sweep cable lengths to find the shortest satisfiable cable constraint.

\paragraph{Cost model.}
% DRAM and CXL have small power impacts~\cite{wang2024designing}, so we focus on purchase cost (CapEx).
% To model the cost overhead of CXL, we assume a hyperscale scenario where a large number of racks adopt CXL.
% While the purchase cost of a smaller CXL pod may be lower than a larger CXL pod, a hyperscaler will have to buy proportionally more of the smaller CXL pods.
% We assume a server purchase cost of \$30K and the estimated cost per \mpd from Figure~\ref{fig:die-area-and-cost}.

% DRAM and CXL have relatively small power overheads~\cite{wang2024designing}, so 
We focus on CapEx.
We model CXL costs in a hyperscale setting and normalize CXL costs by the number of servers. This is because although a smaller CXL pod has lower cost per pod, a hyperscaler must deploy more such pods to serve the same number of servers, resulting in similar per-server CXL cost. We assume a server cost of \$30K~\cite{lenovo-server,supermicro-server} and use the estimated per-\mpd cost from Figure~\ref{fig:die-area-and-cost}.

% We extrapolate slowdown based on latency sensitivity measurements (Figure~\ref{fig:latency:sensitivity}) to determine the fraction of workloads that can run on CXL with \mpds and switches, respectively.
% We consider a 10\% slowdown to be tolerable, which leads us to assume 65\% of DRAM can come from \mpds and 35\% of DRAM can be provisioned with switches, respectively.
% For switches, we assume that we retain some local CXL to fulfill the remaining capacity need.

\begin{comment}
\begin{figure}[t!]
    \centering
    \includegraphics[width=0.4\textwidth]{figures/pod_size_cost_tradeoff.eps}
    % \vspace{-1.4em}
    \caption{Fully-connected CXL pod topologies define a Pareto frontier where cost rises super-linearly with CXL pod size. For medium to large pod sizes, \sysname can achieve much larger pod sizes at equal or lower cost.}
    \label{fig:tradeoff}
\end{figure}
\end{comment}

\subsection{Hardware Prototype}
\label{sec:eval-hw}

% \yuhong{Add hardware experiments to quantify and compare application performance using local
% DDR memory, CXL memory expansion, and CXL memory pooling.}

We briefly describe our current \mpd's characteristics and then evaluate our three-server island prototype.
We show that we can achieve low-latency RPCs, motivate why islands are necessary, and quantify bandwidth gains for island-wide collective communication.
Our prototype server each uses two $\times 8$ CXL ports and a single 100\,Gbit NIC (vs.\ eight ports and 400\,Gbit at production scale), scaling both CXL and Ethernet by the same $4\times$ factor so that CXL-vs.-network comparisons remain representative.

\paragraph{\mpd hardware characteristics.}
% We measure each $\times 8$ CXL link to provide 24.7\,GB/s read-only bandwidth and 22.5\,GB/s write-only bandwidth.
% However, for mixed workloads 1:1 read-write bandwidth increases only to 28.8\,GB/s when a single host is active.
% This is less than expected due to a \mpd firmware issue since the CXL link itself is fully bidirectional.
% When both servers are active, each server's bandwidth maxes out at 22.1\,GB/s.
We measure each $\times8$ CXL link to deliver 24.7\,GiB/s read-only and 22.5\,GiB/s write-only bandwidth.
For mixed 1:1 read-write workloads, total bandwidth increases to only 28.8\,GiB/s.
This is lower than expected given that the CXL link is fully bidirectional, and is due to a \mpd firmware issue.
When both servers are active, per-server bandwidth saturates at 22.1\,GiB/s.

\paragraph{Application slowdown.} Given that MPDs have higher latency than expansion devices (267\,ns vs. 233\,ns, measured with our hardware), we evaluate slowdown across a broad set of cloud workloads: web (Ruby YJIT \cite{ruby-yjit}), key-value stores (YCSB \cite{cooper2010benchmarking} on Redis \cite{redis} and Memcached \cite{memcached}), and databases (TPC-C \cite{tpc-c} on Silo \cite{silo}, and TPC-H \cite{tpc-h} on PostgreSQL \cite{postgres}).

\begin{figure}[t]
    \centering
    \includegraphics[width=0.8\columnwidth]{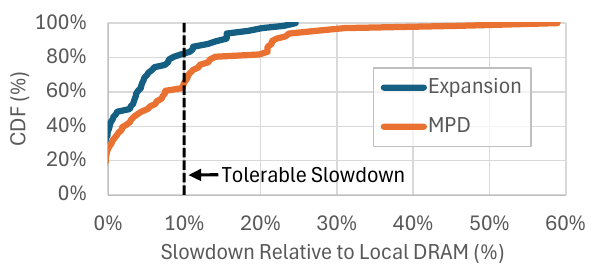}
    \vspace{-0.7em}
    \caption{Cumulative distribution of application slowdown for CXL expansion devices and \mpds relative to local DRAM.}
    \label{fig:mpd_expansion_slowdown}
    \vspace{-1em}
\end{figure}

Figure \ref{fig:mpd_expansion_slowdown} shows that about 65\% of applications incur less than 10\% slowdown with \mpds, consistent with our earlier estimate (§\ref{sec:use-case-pooling}). Note that latency-sensitive workloads that exceed 10\% slowdown with expansion devices degrade further on \mpds; however, such workloads would not be provisioned with CXL memory in practice.

% We believe the cause to be a device-internal bottleneck.
% This bottleneck is likely in the DDR5 memory controller which would need to serve 57.6~GB/s.
% It is possible that our \mpd prototypes operate DIMMs are lower speed, causing a DDR5 bottleneck.
% %We find that this discrepancy is caused by the limited DDR bandwidth per \mpd.
% %For each \mpd, sustaining 28.8~GB/s bandwidth at the two $\times 8$ links requires 57.6~GB/s DDR bandwidth in total.
% %However, due to thermal reasons, our \mpd prototypes operate DIMMs at a lower DDR speed, causing the DDR bandwidth to be the bottleneck.
% We expect that production \mpds will overcome this limitation.

\begin{figure}[t!]
    \begin{subfigure}[b]{0.46\linewidth}
        \centering
        \includegraphics[width=\linewidth]{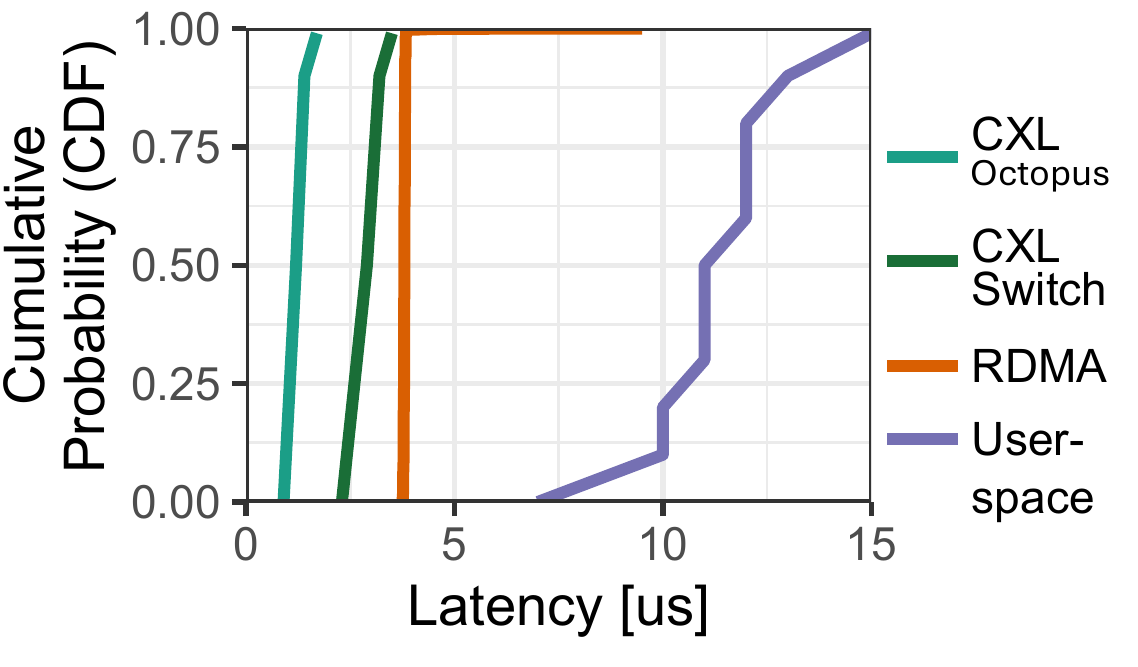}
        \vspace*{-1.7em}
        \caption{64 Bytes.}
        \label{fig:64byterpc}
    \end{subfigure}\hfill
    \begin{subfigure}[b]{0.51\linewidth}
        \centering
        \includegraphics[width=\linewidth]{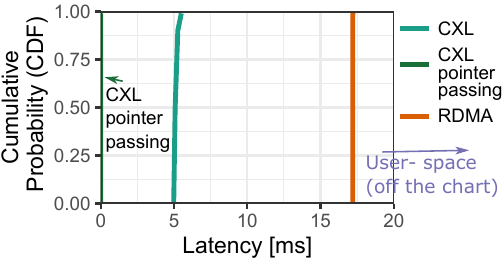}
        \vspace*{-1.7em}
        \caption{100 MB.}
        \label{fig:100mbrpc}
    \end{subfigure}
    \vspace*{-0.5em}
    \caption{Distribution of RPC round-trip latency for small and large messages.
    % CXL's latency is $3.3\times$ and $1.5\times$ lower than RDMA, and $10\times$ lower than user-space networking.
    }
    \vspace*{-0.7em}
    \label{fig:rpc_latency}
\end{figure}

% numbers from fig (a)
%1:       RDMA  3.79996
%2: CXL Acadia  1.20000
%3: CXL Switch  2.90000
%4: User-space 11.00000

% fig (b)
%1:          RDMA 17202.0
%2: CXL full data  5080.0
%3:   CXL pointer     1.2

\paragraph{Small RPCs.}
We measure the round-trip latency of RPC where RPC parameters and return values are both 64\,B.
% With \sysname, given that any pair of servers within an island has a common \mpd, each RPC message is only transmitted through a single \mpd.
Figure~\ref{fig:64byterpc} shows that the median RPC latency within an \sysname island is 1.2\,\us.
A CXL switch has $2.4\times$ higher latency.
RDMA has $3.2\times$ higher latency at 3.8\,\us.
User-space networking has $9.5\times$ higher latency at over 11\,\us.

\begin{figure}[t]
    \centering
    \includegraphics[width=0.8\columnwidth]{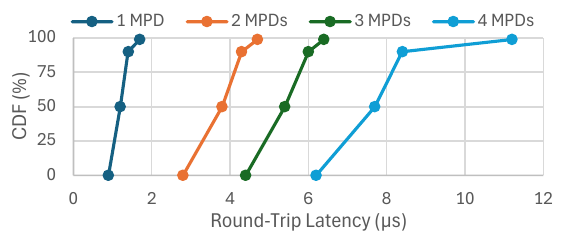}
    \vspace{-0.7em}
    \caption{Round-trip RPC latency with messages being transmitted through a varying number of \mpds. Within an island, passing messages between any pair of servers only go through a single \mpd.}
    \label{fig:rpc-varying-mpds}
    \vspace*{-0.7em}
\end{figure}

We validate \sysname's island concept~(\S\ref{sec:logicaltopo}) by measuring RPC latency when servers have to forward messages because they do not share a \mpd.
For example, expander topologies with 96 servers typically require traversing 3 \mpds.
We measure the round-trip RPC latency with a varying number of intermediate \mpds.
Figure~\ref{fig:rpc-varying-mpds} shows that transmitting a message through two \mpds increases the median latency from 1.2\,\us to 3.8\,\us, comparable to RDMA.

\paragraph{Large RPCs.}
We measure 100\,MB parameters and 64\,B return values.
With \sysname, we can pass RPC parameters by value or by reference.
For the latter we assume that all parameters are already in the \mpd, so there is no need to serialize and copy the parameter.

Figure~\ref{fig:100mbrpc} shows the round-trip RPC latency.
When passing by value, CXL achieves 5.1\,ms median latency. The median latency with RDMA is 3.3$\times$ higher.
When passing by reference, CXL latency matches the 64\,B case, orders of magnitude lower than passing by value.

\paragraph{Broadcast collectives.}
The source server connects directly to the two other servers each via a separate \mpd.
Thus, the source server writes the data to the two \mpds in parallel and both destination servers read the data from the corresponding \mpd in a pipeline while the source server is still writing.
We measure the completion time for broadcasting 32\,GB to two servers at 1.5\,s, confirming the \mpd serves at full bandwidth to both servers in parallel, a $2\times$ speedup over RDMA.
%Each $\times 8$ CXL link at the source host achieves 21.3~GB/s write bandwidth, matching the expected write-only bandwidth of a $\times 8$ link.
%Note that each \mpd provides sufficient DDR bandwidth to serve 21.3~GB/s read and 21.3~GB/s write CXL bandwidth in parallel.

\paragraph{All-gather collectives.}
The CXL links in our three-server island form a cycle, so we use the ring all-gather algorithm.
With 32\,GiB shards per server the all-gather completes in 2.9\,s, achieving 22.1\,GiB/s aggregate bidirectional bandwidth (vs.\ 28.8\,GiB/s expected), limited by the same \mpd firmware bottleneck noted above.

\subsection{Scaled-Up Simulations}
\label{sec:eval-sim}

We use the hardware measurements to simulate scaled-up versions of \sysname with 96 servers.
% Since we do not have the hardware to evaluate large \sysname topologies, we use simulations to demonstrate the benefits of \sysname comparing to other topologies.
% We focus on the cost saving of memory pooling and the completion time of bandwidth-bound communication.

\subsubsection{Memory Pooling Savings}
% We compare the memory pooling savings of \sysname, random topologies (\ie Jellyfish~\cite{singla2012jellyfish}), expander topologies (\ie Xpander~\cite{hoory2006expander}), and BCube~\cite{bcube}.
We show that \sysname achieves almost optimal memory pooling savings among \mpd topologies with $X=8$ and $N=4$.
We also compare \sysname to CXL switches which have better reachability than \mpds.
However, CXL switches also have higher latency and thus can pool a smaller fraction of memory.
By pooling a larger fraction of memory capacity, \sysname is able to match the pooling savings of CXL switches even under very optimistic assumptions on the CXL switch topology.
%demonstrate that although CXL switches have better reachability, \sysname achieves similar pooling savings overall.
%This is because \sysname can pool memory across more VMs, including those that are sensitive to CXL switch latencies.

\paragraph{\sysname vs. other \mpd topologies.}
To show that \sysname achieves near-optimal pooling savings, we compare it against expander topologies (Jellyfish~\cite{singla2012jellyfish}) with varying pod sizes under $X=8$ and $N=4$.
Expander topologies are known to have asymptotically optimal expansion~\cite{singla2012jellyfish,valadarsky2016xpander}.
All topologies have the same \mpd cost per server, since the server-to-\mpd ratio is identical.

% We find that our random topologies behave like expander topologies (\ie Xpander~\cite{hoory2006expander}), and thus report a combined metric.

\begin{figure}
    \centering
    \includegraphics[width=0.4\textwidth]{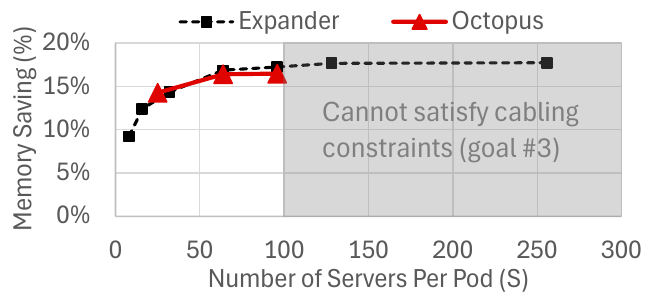}
    \vspace{-0.6em}
    \caption{Average memory pooling savings with \sysname and expander topologies with a varying pod size ($S$).}
    \label{fig:avg-pooling-savings}
    \vspace{-1em}
\end{figure}

Figure~\ref{fig:avg-pooling-savings} shows pooling savings for expander topologies up to 256 servers.
Note that some of these are not physically feasible topologies, as copper cable lengths will not scale beyond two server racks, and roughly 100 severs.
These large and unrealistic \mpd topologies save up to 18\% compared to \sysname-96 which saves 16\%.
% In principle, larger topologies allow better reachability and load balancing among hosts with high memory demands.
The benefits flatten out around 100 servers, consistent with the peak-to-average demand ratios in Figure~\ref{fig:poolingsavings}, validating \sysname's focus on 96 servers with practical copper cabling.

\paragraph{\sysname vs. CXL switches.}
CXL switches are subject to similar limitations as \mpds under fully-connected wiring assumptions.
Specifically, when every CXL switch must connect to every server (as in prior work~\cite{pond:asplos2023}), we cannot connect more than 20 servers in a pod, because we must at least use 10 ports for CXL devices and 2 ports for management~\cite{xconn2024latency}.
When pooling across 20 servers, CXL switches can save 12\% of memory, compared to 16\% for \sysname-96.

We therefore simulate an optimistic CXL switch design to upper-bound its pooling savings. Specifically, we assume switches adopt a sparse topology similar to \sysname, forgo management ports, and connect up to 90 servers. Although such a sparse switch topology would be less efficient than a switch fabric that fully connects servers and devices, we optimistically model it as a global pool with 90 servers. Under this optimistic assumption, the switch topology achieves 16\% memory savings, matching \sysname. This is because it pools only 35\% of total DRAM and saves 46\% of that pooled memory. In contrast, \sysname pools 65\% of total DRAM while saving 25\% of the pooled memory~(\S\ref{sec:use-case-pooling}).
%Therefore, although the switch topology can multiplex memory demands better because of the high switch port count, the high access latency affects the overall pooling savings and has no benefits comparing to \sysname.

\begin{figure}
    \centering
    \includegraphics[width=0.42\textwidth]{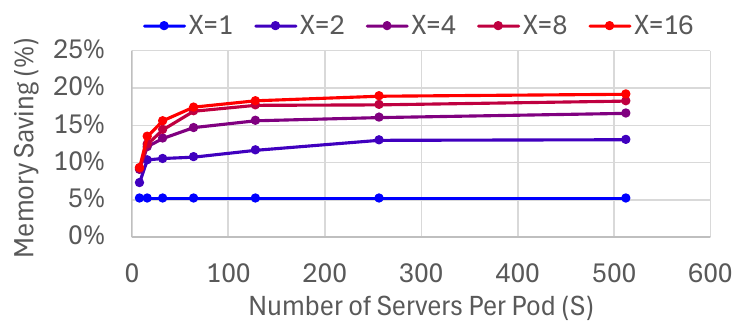}
    \vspace{-0.5em}
    \caption{Average pooling savings of expander \mpd topologies with a varying pod size ($S$) and a varying server port count ($X$).}
    \label{fig:pooling-vary-x}
    \vspace{-0.5em}
\end{figure}

\paragraph{Sensitivity analysis.}
We use expander topologies (Jellyfish~\cite{singla2012jellyfish}) to show how pooling savings vary with pod size ($S$), server port count ($X$), and \mpd port count ($N$).
For reference, \sysname-96 exhibits behavior similar to a expander topology with 64 servers. Figure~\ref{fig:pooling-vary-x} shows memory savings for expander topologies with $S$ ranging from 2 to 512 servers and $X$ from 1 to 16.
Pooling savings generally increase with $X$, with diminishing returns beyond $X=8$.

% To show a wide range of memory savings, we consider random topologies.
% For reference, \sysname-96 behaves similar to a random topology with 64 servers.
% Similarly, a variant of \sysname for $X=4$ would have islands of size 9, giving 36 servers, comparable to a random topology with 32 servers.
% Figure~\ref{fig:pooling-vary-x} shows memory savings for random topologies from 2 to 512 hosts and $X=1$ to $X=16$.
% Generally, memory savings increase with $X$.
% However, $X=16$ provides marginal benefits over $X=8$ .

In terms of \mpd port count ($N$), pooling savings are minimal with $N=2$.
This configuration cannot reach sufficient \mpds even with $X=8$.
Conversely, $N=8$ is far more effective than $N=4$, though no $N=8$ \mpds exist today.

\subsubsection{Bandwidth-Bound Communication}

\begin{figure}[t]
    \centering
    \includegraphics[width=0.42\textwidth]{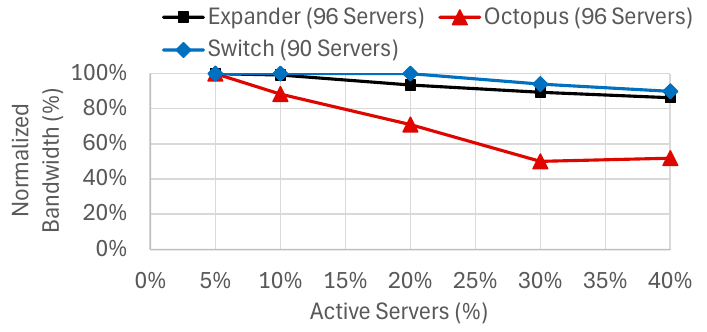}
    \vspace{-0.5em}
    \caption{Average normalized bandwidth under random traffic with a varying number of active servers.}
    \label{fig:bw-sparse}
    \vspace{-1em}
\end{figure}

\sysname provides both low server-to-server communication latency and high bandwidth within each island.
We show that \sysname achieves optimal performance for bandwidth-bound communication within each island.

\paragraph{Single active island.} We simulate the optimal completion time of a uniform all-to-all communication within an island. The optimal completion time is obtained by solving a multi-commodity max flow problem~\cite{maxflow} using linear programming. We assume that only one island is active, which allows the active island to also route traffic through inactive islands.

Our simulations show that all-to-all communication within the active island achieves optimal bandwidth, with each server fully saturating its 8 CXL links (5 intra-island, 3 inter-island), demonstrating that \sysname can leverage unused inter-island link bandwidth.

\paragraph{Random traffic.} However, as \sysname has less inter-island bandwidth, it has lower performance under random communication across the entire pod compared to an expander topology. Figure~\ref{fig:bw-sparse} shows that with 10\% active servers, \sysname has 12\% lower performance.
In contrast, switches achieve higher bandwidth due to their high fanout.

%\paragraph{All-to-all}

\subsubsection{CXL Link Failures}
CXL link failures are the dominant failure factor introduced by CXL memory~\cite{berger2026cxlpractice}.
To study how CXL link failures affect different CXL use cases, we simulate memory pooling savings and bandwidth-bound communication performance with a varying CXL link failure ratio.
For a given failure ratio, we uniformly randomly select CXL links to fail\footnote{Disconnecting CXL devices from a running server (e.g., caused by CXL link failures) is termed \emph{surprise removal of devices} in CXL specifications~\cite{cxl-3.0}.
As of CXL 3.0, surprise removal may cause server faults and reboots.
We therefore assume affected servers have rebooted and can access remaining \mpds via functional links.}.

\begin{figure}
    \centering
    \includegraphics[width=0.4\textwidth]{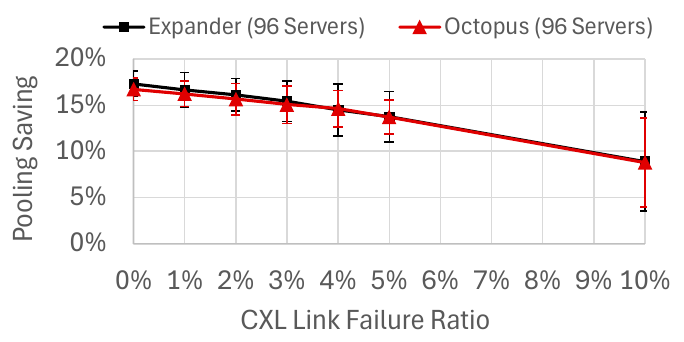}
    \vspace{-0.7em}
    \caption{Average memory pooling savings under a varying CXL link failure ratio. The standard deviation is shown as the error bars.}
    \label{fig:failure-pooling}
    \vspace{-0.4em}
\end{figure}

\paragraph{Pooling savings.} CXL link failures reduce pooling savings because affected servers can access fewer \mpds.
If such a server becomes ``hot,'' its excess memory demand can only be served by the remaining \mpds it connects to, which lowers pooling savings.
% As the link failure ratio increases, hot servers are more likely to concentrate their allocations on fewer MPDs, leading to reduced pooling savings. We will clarify this in the paper.
We compare the memory pooling savings between \sysname and the expander topology with a varying CXL link failure ratio.
Figure~\ref{fig:failure-pooling} shows that the pooling savings of both \sysname and the expander topology degrade gracefully from 17\% to 14\% with 5\% failed CXL links.

\begin{table}[t]
    \centering
    \small
    \begin{tabular}{ccccc}
    \toprule
    Islands & Pod Size & CXL CapEx Cost & Cable Len.\\
    \midrule
    %2  & 2  & 9   & 111\% & 0.6 m\\
    1  & 25  & \$1252 / server & 0.7\,m\\
    4  & 64  & \$1292 / server & 0.9\,m\\
    6  & 96  & \$1548 / server & 1.3\,m\\
    \bottomrule
    \end{tabular}
    \vspace*{-0.5em}
    \caption{\sysname configurations with $X=8$ server ports and $N=4$ \mpd ports.
    The CXL cost includes CXL devices and required cables.
    The cable length is the known minimum required to realize the \sysname topology within a 3-rack configuration.
    Increased cable cost is the main reason for increasing costs of \sysname-96.}
    \label{tab:configex}
    \vspace*{-0.8em}
\end{table}

\begin{table}[t]
    \centering
    \small
    \begin{tabular}{ccccc}
    \toprule
    Topology & Pod Size & CXL CapEx & Mem Saving\\
    \midrule
    Expansion & \textbackslash & \$800 / server & \textbackslash \\
    \sysname  & 96  & \$1548 / server & 16\% \\
    Switch  & 90  & \$3460 / server & 16\% \\
    \bottomrule
    \end{tabular}
    \vspace*{-0.7em}
    \caption{Comparison of CXL device CapEx and memory pooling cost savings of \sysname and CXL switches.}
    \label{tab:cost-cmp}
    % \vspace{-0.5em}
\end{table}

\begin{table}[th!]
    \centering
    \small
    \begin{tabular}{lcccc}
    \toprule
    Power factor &
    $1.00$ &
    $1.25$ &
    $1.50$ &
    $2.00$ \\
    \midrule
    Switch CapEx {(\$/server)} &
    \$2969 &
    \$3589 &
    \$4613 &
    \$9487 \\
    Server CapEx & +1.7\% & +3.7\% & +7.1\% & +22.9\% \\
    \bottomrule
    \end{tabular}
    \vspace*{-0.8em}
    \caption{Switch cost sensitivity under a power-law die-area model.}
    \label{tab:power-model}
    \vspace*{-1em}
\end{table}

\paragraph{Communication performance.}
% Figure~\ref{fig:failure-comm-octopus} shows the normalized communication bandwidth of \sysname under random traffic as the CXL link failure ratio increases.
We also simulate communication under random traffic with CXL link failures.
With 5\% link failures, performance degrades by 5\%--12\%, indicating \sysname's path diversity helps sustain performance.

% \begin{figure}[!t]
%     \centering
%     \includegraphics[width=0.43\textwidth]{figures/failure-comm-octopus.pdf}
%     \vspace{-0.5em}
%     \caption{Normalized communication bandwidth as a function of CXL link failure ratio with a varying number of active servers.}
%     \label{fig:failure-comm-octopus}
% \end{figure}

% \yuhong{Others to discuss: CXL switch failure simulation; using multiple physical links to run in degraded mode; interrupt rate}

\subsection{Physical Layout Validation}
\label{sec:eval-phy}

%\begin{figure}
%    \centering
%    \includegraphics[clip,width=0.44\textwidth]{figures/RackDesign_9.png}
%    \caption{A 3-rack implementation of \sysname with 9 hosts and 36 2-port \mpds, with corresponding logical topology shown in Figure~\ref{fig:acadia9}. This layout requires 0.6 m cables. Logical host and device ids are shown in the slots they occupy.}
%    \label{fig:rack_design9}
%\end{figure}

To find physical placements for \sysname topologies within a 3-rack system~(\S\ref{sec:physicallayout}), we solve the physical mapping SAT problem for various cable lengths and determine the length required to realize each topology in Table~\ref{tab:octopus-topo}. 
The cable lengths for which a solution was found within 48 hours of wall-clock time are documented in Table~\ref{tab:configex}.
All three \sysname topologies can be realized with cable lengths of 1.3\,m or less.

\subsection{Cost Comparison}
\label{sec:cost-cmp}
\paragraph{CXL CapEx.} Table~\ref{tab:cost-cmp} summarizes the CXL device CapEx and memory pooling cost savings of the 96-server \sysname and the 90-server CXL switch topology. Although \sysname and the CXL switch topology achieve the same memory pooling savings, the switch topology’s device cost is more than twice that of \sysname.
\sysname's cost is 5\% of server CapEx vs.\ 12\% for switches.

\paragraph{Server CapEx.} Combining device CapEx with memory pooling savings, \sysname reduces overall server CapEx by 3.0\% compared to servers without CXL.
This reduction may appear small at first, but is significant at hyperscale.
Moreover, this estimate excludes additional savings from data sharing and communication use cases, which can further reduce CPU core requirements by lowering RPC overheads.

% Combining the device CapEx and memory pooling savings shows that \sysname reduces the overall server CapEx by 3.0\% comparing to servers without CXL.
% This may appear small at first, but this is significant at hyperscale.
% Also note that we have not accounted for any savings from data sharing and communication use cases.
% These other use cases may be able to reduce required CPU core counts, by reducing RPC overheads.

The overall server CapEx of the switch topology is 3.3\% \emph{higher} than the baseline without CXL.
Due to the expensive price of switch and the many cables required, the CXL switch topology costs more than it saves in terms of DRAM CapEx.

So far, we have assumed a server without CXL as a baseline.
Since CXL memory expansion is seeing increasing adoption~\cite{azure-cxl-m-series}, we consider a baseline where the server includes CXL memory expansion.
In this case, \sysname reduces the server CapEx by 5.4\%.
The key is that \sysname's \mpds cost little more than existing expansion devices, whereas the switch topology still increases CapEx by 0.6\%.

\paragraph{Sensitivity analysis.} We also evaluate an alternative switch cost model in which die cost scales as a power law with die area, reflecting non-linear yield effects in semiconductor manufacturing. We vary the power factor starting from 1.0 (linear scaling). Table~\ref{tab:power-model} shows that even under the optimistic linear model, server CapEx still increases by 1.7\%.

\section{Discussion}\label{sec:discussion}

\paragraph{Limitations.}
In terms of communication, \sysname is optimized for low-latency and high-bandwidth communication \emph{within} an island.
Communication between servers in \emph{different} islands may traverse two \mpd hops, incurring higher latency and lower bandwidth compared to fully-connected \mpd or switched topologies.
In addition, broadcast-style collectives (\eg all-gather over shared CXL memory~\cite{ahn2024mpi,cxl-mem-mpi,cmpi}) also favor full connectivity, as a sender can write once and all receivers read from the same \mpd; in \sysname, such patterns require multi-hop reads or replication across \mpds.

For memory pooling, \sysname may have suboptimal savings when server demands are extremely skewed.
For example, if a single server needs to allocate nearly all CXL memory, only a fully-connected (\mpd or switch) topology can serve this extreme skew; \sysname's per-server \mpd reachability is bounded by the number of \mpds a server connects to.

Finally, \sysname's irregular cabling may be harder to manage than a single-switch or fully-connected \mpd pod~\cite{singla2012jellyfish,valadarsky2016xpander}, though the comparison is less clear at sub-rack scale.

% \sysname is optimized for low-latency access and high bandwidth \emph{within} an island. Communication between servers in \emph{different} islands may traverse two \mpd hops, which is adequate for pooling but provides lower bandwidth than a fully connected or switched topology for randomly chosen host pairs.
% Similarly, if a single host needs to allocate nearly all pod memory, only a fully connected (\mpd or switch) topology can serve this extreme skew; \sysname's per-host reachability is bounded by the number of \mpds a host connects to.
% Broadcast-style collectives (\eg AllGather over shared memory~\cite{ahn2024mpi}) also favor full connectivity, as a sender can write once and all receivers read from the same \mpd; in \sysname, such patterns require multi-hop reads or replication across \mpds.
% Finally, \sysname's irregular cabling may be harder to manage than a single-switch or fully connected pod~\cite{singla2012jellyfish,valadarsky2016xpander}, though the comparison is less clear at sub-rack scale.

\paragraph{Port count changes.}
\sysname assumes $X=8$ CXL ports per server and $N=4$ ports per \mpd, based on existing Intel Xeon 6 and AMD 5th Gen EPYC deployments~\cite{intel-xeon-6,amd-epyc-5}.
CXL 4.0 over PCIe~6.0 will double per-lane bandwidth, making narrower links ($\times 4$ or even $\times 2$) viable, so $X=8$ independent $\times 4$ links on 32 total lanes is realistic and \mpds can scale to $N \geq 4$ within today's form factors.
\sysname's topologies are specific to $X$ and $N$; the split between island-specific ports ($X_i$) and cross-island ports ($X - X_i$) must be re-optimized for each configuration, which we leave to future work.

\paragraph{CXL switch topologies and future interconnects.}
CXL 3.0 enables non-tree switch topologies and direct switch-to-switch links~\cite{cxl-3.0}, but each switch hop adds ${\approx}\,220$\,ns of latency, making multi-level switching unattractive for latency-sensitive workloads.
A promising middle ground is to combine \mpd-based \sysname islands with a small switch fabric for global reachability.
A transition from copper to optical CXL links could extend reach but does not eliminate the need for topology design as long as switches remain electrical.

\paragraph{Security.}
Within a single server, CXL memory isolation relies on the OS or hypervisor page tables, exactly as with local DRAM.
Across servers sharing a CXL device, CXL~2.x provides no inter-server access control; pooling and sharing must also rely on hypervisor page tables.
CXL~3.x Dynamic Capacity Devices (DCD)~\cite{ha2023dynamic,cxl-3.0} adds hardware-enforced, per-server access control for shared regions, enabling fine-grained multi-tenant isolation.
\sysname works with both: static \mpd partitioning under CXL~2.x, and on-demand secure sharing via DCD under CXL~3.x.

\paragraph{Memory allocation.} Memory allocation for pooling in \sysname is challenging: greedily allocating from \mpds shared with neighbors may cause contention when those servers later become hot. Whether per-server demand prediction or limited memory migration can improve pooling efficiency remains an open problem.

In addition, bandwidth-sensitive workloads may require software interleaving across multiple \mpds, making pooled memory allocation more challenging when bandwidth constraints are considered.

\paragraph{Multi-socket servers.} \sysname supports multi-socket servers in two ways: treating each server as a single node, or treating each socket as a separate node.
This choice again trades off pod size and latency.
Treating an entire server as one node can enable larger pods, but communication may traverse inter-socket links (\eg UPI), increasing latency.

\section{Related Work}

\paragraph{CXL pod topologies.}
\sysname is, to our knowledge, the first work to compare sparsely connected CXL pod \emph{topologies} and layouts rather than assuming one wiring pattern.
Most prior work and industry decks adopt fully connected \mpd pods or switch fabrics and study device properties, not topology tradeoffs across pooling and communication~\cite{wagh2021cxl,seagate2024cxl,arm2024cxl,ha2023dynamic,pond:asplos2023,mackey2024cxl,hyatt2023quest}. 
Typical designs connect every CPU to the same $X$ \mpds so that $N$-ported \mpds yield pods of size $N$; this scales poorly in port count~\cite{wagh2021cxl,seagate2024cxl,arm2024cxl,ha2023dynamic,pond:asplos2023,mackey2024cxl,hyatt2023quest}.

\paragraph{Scale-up “islands.”}
Large shared-memory systems (HPE Superdome Flex, SGI UV) compose high-bandwidth islands with longer inter-island paths; \sysname islands mirror this: dense local sharing on a common \mpd with sparser inter-island links, aligning with MDC/OpenFAM-style locality-aware programming~\cite{hpe_superdome_flex_arch_ras_2021,sgi_uv300_ras_2016,keeton_mdc_fast17,singhal_openfam_2022,keeton_openfam_api_2019}.

\paragraph{Interconnection evolution.}
Point-to-point device fabrics and switches have traded places across decades: trees to high-radix, low-diameter designs (flattened butterfly, Slim/Xpander), and Clos-like CXL 3.0 fabrics~\cite{bjerregaard2006survey,dally1990performance,leiserson1985fat,kim2007flattened,kim2005microarchitecture,kim2008technology,besta2014slim,lakhotia2022polarfly,blach2024high,cxl2topology,cxl3fabric}. \sysname targets near-term copper-first deployments with sparse MPD bipartite graphs with one-hop sharing while expanding pooling reach.
Server-forwarded schemes like BCube~\cite{bcube} are less attractive under CXL latency.

\section{Conclusion}

\sysname challenges the assumption that effective CXL pods require fully-connected topologies with large pooling devices, showing that sparsely-connected topologies can be cost-efficient for pooling and low-latency communication.
% We find that \sysname pods provide benefits to memory-intensive cloud and database workloads.
\sysname offers a practical blueprint for CXL memory pooling that balances cost, scalability, and performance.

\section*{Acknowledgments}

We thank our shepherd, Alex C. Snoeren, and the anonymous reviewers for their helpful comments. We also thank Midhul Vuppalapati for his feedback and contributions to the memory pooling analysis.

%abbrv
\bibliographystyle{abbrv}
\bibliography{references}

\newpage
\appendix
\section{Appendix}

\subsection{Expansion and Pooling Benefits}
\label{app:expansion}

The expansion $e_k$ of an \mpd topology directly yields a lower bound on peak \mpd usage within a pod:
\begin{theorem}
Let $D_k$ denote the maximum aggregate memory demand of any subset of $k$ servers in a pod.
Assume a worst-case scenario in which, for each $k$, a subset of $k$ servers $U \subseteq S$ that attains $D_k$ also connects to only $e_k$ distinct \mpds.
Then the peak memory usage across all \mpds, denoted $L^{\star}$, satisfies
\[
L^{\star}\ge\max_{1\le k \le |S|}\frac{D_k}{e_k},
\]
where $|S|$ is the number of servers in the pod.
\end{theorem}

\begin{proof}
Fix $k$ and let $U$ be a subset of $k$ servers that attains $D_k$ and, by the assumption, connects to $e_k$ distinct \mpds.
All demand from servers in $U$ must be served by these $e_k$ \mpds, so at least one of them must carry load at least $D_k/e_k$.
Therefore $L^{\star} \ge D_k/e_k$ for any $k$.
Taking the maximum over $k$ completes the proof.
\end{proof}

\end{document}